\documentclass[10pt]{article}

\usepackage[utf8]{inputenc}
\usepackage{enumerate}
\usepackage{units}

\usepackage{amsthm}

\usepackage{amsmath}
\usepackage{amsfonts}
\usepackage{amssymb}
\usepackage{amscd}
\usepackage{latexsym}

\usepackage{mathrsfs}

\usepackage{xy}

\usepackage[switch,mathlines]{lineno}

\def\X{\mathcal X}
\def\C{\mathbb{C}}

\def\N{\mathbb{N}}
\def\R{\mathbb{R}}

\def\A{{\mathcal A}}
\def\B{{\mathcal B}}
\def\CC{C}

\def\F{\mathcal F}
\def\H{\mathcal H}
\def\K{\mathcal K}
\def\M{\mathcal M}
\def\Q{\mathcal Q}

\def\S{\mathcal S}

\def\U{\mathcal U}
\def\CCC{{\mathfrak C}}
\def\FF{{\mathfrak F}}

\def\KK{{\mathfrak K}}
\def\r{{\mathfrak r}}

\def\Rep{\mathfrak{Rep}}
\def\dia{\diamond^{\hbox{\it \tiny B}}}
\def\de{\mathrm{d}}

\def\im{\mathop{\mathsf{Im}}\nolimits} 
\def\I{{\rm 1\kern-.26em I}}
\def\SA{S_{\!\A}}
\def\FA{F_{\!\A}}
\def\oB{\omega^{\hbox{\it \tiny B}}}
\def\Ob{\Omega_{\hbox{\it \tiny B}}}
\def\ob{\omega_{\hbox{\it \tiny B}}}

\def\gb{\Gamma_{\!\hbox{\it \tiny B}}}
\def\sb{\Sigma_{\hbox{\it \tiny B}}}

\def\MB{{\mathcal M}^{\hbox{\it \tiny B}}}
\def\var{\kappa}
\def\z{\mathsf{z}}

\def\CBA{{\mathfrak C}^{B}_{\!\A}}
\def\BBA{{\mathfrak B}^{B}_{\!\!\A}}

\def\T{\X^*;\mathcal A^\infty}
\def\sh{\;\!\sharp^{\hbox{\it \tiny B}}}
\def\AB{\mathfrak A^{\hbox{\it \tiny B}}(\Xi)}
\def\MBA{{\mathfrak M}^{B}_{\!\!\A}}

\def\b{\mathfrak{b}}
\def\a{\mathfrak{a}}

\def\1{\mathfrak{1}}
\def\0{\mathfrak{0}}
\def\G{\mathcal{G}}
 
\def\p{\mathfrak{p}}

\def\<{\langle}
\def\>{\rangle}

\def\Op{\mathfrak{Op}}

\providecommand{\CCC}{\mathfrak{C}}
\def\SS{\mathfrak{S}}

\providecommand{\abs}[1]{\left | #1 \right |}
 \providecommand{\babs}[1]{\bigl
| #1 \bigr |} 

 \providecommand{\bnorm}[1]{\bigl
\Vert #1 \bigr \Vert}

\providecommand{\expval}[1]{\left \langle #1 \right \rangle}

 \providecommand{\noverk}[2]{\left (
\begin{matrix}
    #1 \\
    #2 \\
\end{matrix}
\right )}

\newtheorem{lemma}{Lemma}[section]
\newtheorem{corollary}[lemma]{Corollary}
\newtheorem{theorem}[lemma]{Theorem}
\newtheorem{proposition}[lemma]{Proposition}
\newtheorem{definition}[lemma]{Definition}

\newtheoremstyle{remark_rm}{}{}{\upshape}%
{}{\bf}{.}{ }{}

\theoremstyle{remark_rm}
\newtheorem{remark}[lemma]{Remark}
\numberwithin{equation}{section}

\begin{document}

\title{Magnetic pseudodifferential operators with coefficients in $C^*$-algebras}

\date{\today}

\author{M. Lein$^1$, M. M\u antoiu$^2$ and S. Richard$^3\footnote{On leave from Universit\'e de Lyon;
Universit\'e Lyon 1; CNRS, UMR5208, Institut Camille Jordan, 43 blvd du 11
novembre 1918, F-69622 Villeurbanne-Cedex, France.
 }\ $}
\date{\small}
\maketitle \vspace{-1cm}

\begin{quote}
\emph{
\begin{itemize}
\item[$^1$] Technische Universit\"at M\"unchen, Zentrum Mathematik, Department m5,
85747 Garching near Munich, Germany
\item[$^2$] Departamento de Matem\'aticas, Universidad de Chile, Las Palmeras 3425, Casilla 653,
Santiago, Chile
\item[$^3$] Department of Pure Mathematics and Mathematical Statistics,
Centre for Mathematical Sciences, University of Cambridge,
Cambridge, CB3 0WB, United Kingdom
\item[] \emph{E-mails:} lein@ma.tum.de, Marius.Mantoiu@imar.ro, sr510@cam.ac.uk
\end{itemize}
  }
\end{quote}

\begin{abstract}
In previous articles, a magnetic pseudodifferential calculus and a family of $C^*$-algebras associated
with twisted dynamical systems were introduced and the connections
between them have been established. We extend this formalism to symbol
classes of H\"ormander type with an $x$-behavior modelized by an
abelian $C^*$-algebra. Some of these classes generate $C^*$-algebras
associated with the twisted dynamical system. We show the relevance of
these classes to the spectral analysis of pseudodifferential operators with anisotropic symbols and magnetic fields.
\end{abstract}

\section{Introduction}

In previous works \cite{MP1,MPR1,IMP1} a twisted form
of the usual Weyl calculus and of the corresponding crossed product $C^*$-algebras has been introduced.
We refer to \cite{Lu,Mu,KO1,KO2,Le} for related works.
The twisting is defined by a $2$-cocycle
on the group $\R^n$ with values in the unitary group of a function algebra.
The calculus is meant to model the family of observables of a
physical system consisting in a spin-less particle moving in the
euclidean space $\R^n$ under the influence of a variable magnetic field $B$.
It goes without saying that the standard theory is recovered for $B=0$.
The $2$-cocycle is defined by fluxes of the magnetic field through
simplexes and it corresponds to a modification of the canonical
symplectic structure of the phase space $\R^{2n}$ by a magnetic contribution.
Actually the modified symplectic form defines a new Poisson algebra
structure on the smooth classical observables on $\R^{2n}$ and it was
shown in \cite{MP2} that the twisted form of the Weyl calculus
constitutes a strict deformation quantization in the sense of Rieffel
\cite{La,Rie} of the usual Poisson algebra.

A basic requirement for a magnetic pseudodifferential theory is gauge covariance.
A magnetic field $B$ being a closed $2$-form in $\R^n$, it can be generated in
many equivalent ways by derivatives of $1$-forms, traditionally named vector potentials.
These vector potentials are involved in the process of prescribing
operators (intended to represent quantum observables) to classical functions defined on the phase space.
Different equivalent choices should lead to unitarily equivalent
operators, and this is indeed the case for our formalism, see Section \ref{mpsd}, in contrast to previous wrong attempts.

Most often the usual pseudodifferential calculus is studied in the
framework of the H\"ormander symbol classes $S^m_{\rho,\delta}(\R^{2n})$.
The necessary magnetic adaptations, nontrivial because of the bad
behavior of the derivatives of the magnetic flux, were performed in \cite{IMP1}.
Among others, the following results were obtained:
good composition properties, asymptotic developments, an extension of
the Calderon-Vaillancourt result on $L^2$-boundedness, selfadjointness
of elliptic operators on magnetic Sobolev spaces and positivity properties.
A short recall of the magnetic pseudodifferential theory may be found in Sections \ref{veil} and \ref{mpsd}.

Beside the order of a pseudodifferential operator defined by a symbol
$f$, another useful information is the properties of the coefficients,
{\it i.e}.~the behavior of the function $x\mapsto f(x,\xi)$ at fixed $\xi$.
One possible way to take them into account is to confine them to some
abelian $C^*$-algebra $\A$ of functions on $\R^n$.
In the framework of the standard calculus this was performed in a
variety of situations, with a special emphasis on almost periodic
functions, and with various purposes.
We cite for example \cite{Ba1,Ba2,CMS,Co,Sh1}.
In Section \ref{secintro} we are going to investigate the corresponding magnetic case, insisting on composition properties.
We also use this occasion to improve previous results on asymptotic developments.

As soon as the symbol spaces with coefficients in $\A$ are
shown to possess good properties, they can be used to define
non-commutative $C^*$-algebras composed of distributions in phase-space.
Such algebras are investigated in Section \ref{acrp1}.
Then, a partial Fourier transformation makes the connection with the
approach of \cite{MPR1} recalled in Section \ref{calll}.
In that reference, relying on general constructions of \cite{PR1,PR2},
magnetic $C^*$-algebras were introduced in relation with twisted $C^*$-dynamical systems.
These $C^*$-algebras are called twisted crossed products and can be
defined by a universal property with respect to covariant representations.
And once again the $2$-cocycle obtained by the flux of the magnetic
field is the main relevant object, defining both the
twisted action and the algebraico-topological structure of the non-commutative $C^*$-algebras.
Through various representations, these algebras will become concrete $C^*$-algebras
of magnetic pseudodifferential operators in natural Hilbert spaces.

The non-commutative $C^*$-algebras composed of distributions in
phase-space can be generated by $\A$-valued symbols of strictly
negative orders, as shown in Section \ref{acrp1}.
But having in mind applications to the spectral analysis of unbounded
operators, we undertake in Section \ref{mmrr} the task to relate positive order symbols to these algebras.
The key ingredient for that purpose is to understand inversion with respect to the magnetic composition law, or equivalently, to understand inversion of magnetic pseudodifferential operators.
This is the subject of Section \ref{versi}. Among others we show that the inverse of a real elliptic symbol of
order $m>0$ with coefficients in $\A$ is a symbol of order $-m$, also with coefficients in $\A$.
Combined with results of the previous section, this implies that such a
symbol defines an affiliated observable, meaning that its $C_0$-functional calculus is
contained in the twisted crossed product $C^*$-algebra.
We also obtain that the $\A$-valued symbols of order $0$ form a
$\Psi^*$-algebra, and in particular that this algebra is spectrally invariant.

These results on inversion rely at a crucial step on a theorem from \cite{IMP2}.
This theorem, which characterizes magnetic pseudodifferential operators of
suitable classes by their behaviors under successive commutators, is an extension of classical results of Beals and Bony.
Since \cite{IMP2} is not yet published and since the main result on inversion will be used in the following section, we give
an independent proof for the affiliation in an Appendix, extending the
approach of \cite{MPR2}.

Our main motivation was spectral analysis, and the last section is devoted to this subject.
Even for the simplest magnetic differential operators the
determination of its spectrum involves a rather high degree of complexity.
The main reason is that even though the magnetic field is the relevant
physical object, the operators are defined by a vector potential.
Such vector potentials are not unique and one problem is to show the
independence of the result of a particular choice.
Another difficulty is that usually any vector potential defining a
magnetic field will be ill-behaved compared to the magnetic field itself.
For example, bounded magnetic fields might not admit any bounded
vector potential, certain periodic magnetic fields are only defined
by non-periodic vector potentials, etc.
And on the top of all that, general pseudodifferential operators with
magnetic fields were even not correctly defined a couple of years ago.

So Section \ref{secspectral} is devoted to spectral theory.
We investigate the essential spectrum of magnetic pseudo\-differential
operators affiliated to the non-commutative algebras mentioned before.
The key of this approach is the use of the structure of the twisted
crossed products; see \cite{GI1,GI3,GI2,M1,Ric} for related approaches in
the absence of magnetic field, and also \cite{HM,LS} for a description of the
essential spectrum for certain classes of magnetic fields.
We will show how to find information on the essential spectrum in the quasi-orbit structure of the
Gelfand spectrum of the $C^*$-algebra $\A$.

In particular, this allows us to express in Section \ref{esentz} the essential
spectrum of any elliptic magnetic pseudodifferential operators defined by a symbol of
positive order and with coefficient in $\A$ in terms of simpler operators
that are defined on the quasi-orbits at infinity.
For example, our approach covers generalized Schr\"odinger operators of the
form $h(-i\partial-A)+V$, with $h$ a real elliptic symbol of positive order,
and with $V$ and the components of the magnetic field $B$ in some smooth
subalgebra of $\A$. But more generally, our approach works for any operator of the form $f(-i\partial
-A,X)$, once suitably defined, for $f$ a real and elliptic symbol of positive order with coefficients in $\A$.
We stress that there is no condition on $A$, only the components of
the magnetic fields have to satisfy some smoothness conditions and have to belong to $\A$.
We also emphasize that even in the degenerate case $B=0$, we have not been able to locate in the literature a procedure for the calculation
of the essential spectrum of such general pseudodifferential operators with coefficients in some abelian $C^*$-algebra $\A$.

It is rather obvious that the formalism and techniques of this article can be further developed and extended.
More general twisted actions can be taken into account, {\it cf.}~\cite{Rie} for the untwisted case.
This would open the way towards applications to random magnetic operators,
which is the topic of a forthcoming article. Our approach might also be relevant for index theory.
On the other hand, the groupoid setting has shown its role in
pseudodifferential theory, in $C^*$-algebraic spectral analysis and in quantization; we cite for example
\cite{La,LMN,LN,NWX}. Groupoids with $2$-cocycles and associated $C^*$-algebras
are available \cite{Re}, but they are still largely ignored in connection with applications. Extending the pseudodifferential
calculus and the spectral theory to such a framework would be an interesting topic.

{\bf Acknowledgements:} M. M\u antoiu is partially supported by
{\em N{\'u}cleo Cient{\'\i}fico ICM} P07-027-F {\em ``Mathematical
Theory of Quantum and Classical Magnetic Systems''} and by the Chilean Science Fundation 
{\it Fondecyt} under the Grant 1085162. 
S. Richard is supported by the Swiss National Science Foundation.
Discussions with V. Iftimie and R. Purice were one of our sources of inspiration.

\section{Pseudodifferential theory}\label{secintro}

\subsection{The magnetic Moyal algebra}\label{veil}

We recall the structure and the basic properties of the magnetic Weyl calculus
in a variable magnetic field. The main references are
\cite{MP1} and \cite{IMP1}, which contain further details and technical developments.

Let $\X:=\R^n$ and let us denote by $\X^*$ the dual space of $\X$; the duality is given by
$\X\times\X^*\ni(x,\xi)\mapsto x\cdot\xi$. The Lebesgue measures on
$\X$ and $\X^*$ are normalized in such a way that the Fourier
transform $(\F f)(\xi)=\int_\X \de x\,e^{i x \cdot \xi}f(x)$ induces a unitary map from $L^2(\X)$ to
$L^2(\X^*)$. The phase space is $\Xi:=T^*\X\equiv\X\times\X^*$ and the notations
$X=(x,\xi)$, $Y=(y,\eta)$ and $Z=(z,\zeta)$ will be systematically used
for its points. If no magnetic field is present, the standard symplectic form on $\Xi$ is given by
\begin{equation}\label{simp}
\sigma(X,Y) \equiv \sigma \bigl ((x,\xi),(y,\eta) \bigr ) :=
y\cdot \xi-x\cdot\eta\ .
\end{equation}

The magnetic field is described by a closed $2$-form $B$ on $\X$. In the standard
coordinates system on $\X$ it is represented by a function taking
real and antisymmetric matrix values $\{B_{jk}\}$, with $j,k\in \{1,\dots,n\}$,
and verifying the relation $\partial_j B_{kl}+ \partial_k B_{lj}+\partial_l B_{jk}=0$.
We shall always assume that the components of the magnetic fields are
smooth functions, and additional requirements will be imposed when needed.

Classically, the effect of $B$ is to change the geometry of phase space, by adding an extra term to (\ref{simp}):
$\sigma^{\hbox{\it \tiny B}}:=\sigma+\pi^*B,$ where $\pi^*$ is the pull-back associated to the
cotangent bundle projection $\pi:\Xi\rightarrow\X$. In coordinates one has
\begin{equation*}
(\sigma^{\hbox{\it \tiny B}})_{(Z)}(X,Y) = y\cdot \xi - x \cdot \eta +
B(z)(x,y) = \sum_{j=1}^n (y_j \xi_j - x_j \eta_j ) + \sum_{j,k=1}^n
B_{jk}(z) \, x_j y_k\ .
\end{equation*}
Associated with this new symplectic form, one assigns the Poisson
bracket acting on elements $f,g\in C^\infty(\Xi)$:
\begin{equation*}
\{ f , g \}^{\hbox{\it \tiny B}} = \sum_{j=1}^n \bigl ( \partial_{\xi_j} f \,
\partial_{x_j} g - \partial_{\xi_j} g \, \partial_{x_j} f \bigr ) +
\sum_{j,k=1}^n B_{jk} \,\partial_{\xi_j} f \, \partial_{\xi_k} g\ .
\end{equation*}

It is a standard fact that $C^\infty(\Xi;\R)$ endowed with
$\{\cdot,\cdot\}^{\hbox{\it \tiny B}}$ and with the pointwise
multiplication is \emph{a Poisson algebra}, {\it
  i.e.}~$C^\infty(\Xi;\R)$ is a real abelian algebra and
$\{\cdot,\cdot\}^{\hbox{\it \tiny B}}:C^\infty(\Xi;\R) \times
C^\infty(\Xi;\R) \to C^\infty(\Xi;\R)$ is an antisymmetric bilinear
composition law that satisfies the Jacobi identity and is a derivation
with respect to the usual product.

In the quantum picture, the magnetic field $B$ comes into play in
defining a new composition law in terms of its fluxes through
triangles. For $x,y,z \in \X$, let $\<x,y,z\>$ denote the triangle in
$\X$ of vertices $x$, $y$ and $z$ and let us set
\begin{equation*}
\Gamma^B(\<x,y,z\>) := \int_{\<x,y,z\>} B
\end{equation*}
for the flux of $B$ through this triangle (integration of a $2$-form
through a $2$-simplex). With this notation, one defines
\emph{the Moyal product} by the formula
\begin{equation}\label{produitMoyal}
( f \sh g )(X):= 4^n \int_{\Xi} \de Y \int_{\Xi}
\de Z \, e^{-2i\sigma(Y,Z)} \, e^{-i\Gamma^B(\<x-y-z,x+y-z,x-y+z\>)} \, f(X-Y) \, g(X-Z)
\end{equation}
for $f,g:\Xi\rightarrow \C$. For $B=0$ it coincides with the
Weyl composition of symbols in pseudodifferential theory. The composition law $\sh$ provides an
intrinsic algebraic structure underlying the multiplication of the {\it
magnetic pseudodifferential operators} that are going to be defined below.

The integrals defining $f\sh g $ are absolutely convergent only for a
restricted class of symbols. In order to deal with more general
distributions, an extension by duality was proposed in \cite{MP2} under an
additional condition on the magnetic field. So let us assume
that the components of the magnetic field are
$\CC^\infty_{\hbox{\tiny \rm pol}}(\X)$-functions, {\it i.e.}~they are
indefinitely derivable and each derivative is polynomially bounded,
and let $\S(\Xi)$ denote the Schwartz space on $\Xi$. Its dual is
denoted by $\S'(\Xi)$. Then $\S(\Xi)$ is stable under $\sh$, and the
product can be extended to maps $\S(\Xi) \times \S'(\Xi) \to
\S'(\Xi)$ and $\S'(\Xi) \times \S(\Xi) \to \S'(\Xi)$. Denoting
by $\MB(\Xi)$ the largest subspace of $\S'(\Xi)$ for which
$\S(\Xi) \sh \MB(\Xi) \subset \S(\Xi)$ and
$\MB(\Xi) \sh \S(\Xi) \subset \S(\Xi)$,
it can be shown that $\MB(\Xi)$ is an involutive algebra under $\sh$
and under the involution ${}^{\sh}$ obtained by complex conjugation, for which one also has
$\S'(\Xi) \sh \MB(\Xi) \subset \S'(\Xi)$
and $\MB(\Xi) \sh \S'(\Xi) \subset \S'(\Xi)$.

The \emph{Moyal algebra} $\MB(\Xi)$ is quite a large class of
distributions, containing the Fourier transform of all  bounded
measures on $\Xi$ as well as the class $C^\infty_{\rm{pol,u}}(\Xi)$ of all smooth
functions on $\Xi$ having polynomial growth at infinity uniformly in
all the derivatives. In addition, if we assume that all the
derivatives of the functions $B_{jk}$ are bounded, the H\"ormander
classes of symbols $S^m_{\rho,\delta}(\Xi)$ belong to $\MB(\Xi)$ and 
compose in the usual way under $\sh$:
\begin{equation}\label{sifil}
S^{m_1}_{\rho,\delta} (\Xi)\; \sh\; S^{m_2}_{\rho,\delta}(\Xi) \subset
S^{m_1 + m_2}_{\rho,\delta}(\Xi)\ ,
\end{equation}
for $m_1,m_2\in\R$ and $0\leq \delta<\rho\leq 1$ or $\rho=\delta=0$.
Here we have used the following standard definition:
\begin{definition}
The space $S^m_{\rho,\delta}(\Xi)$ of \emph{symbols of order $m$ and of type $(\rho,\delta)$} is
\begin{equation*}
\Big\{f\in C^\infty(\Xi)\mid \forall \alpha,a\in
\N^n,\exists \  C_{\alpha a}<\infty\ \hbox{\rm s.t.}~|
(\partial^a_x\partial^\alpha_\xi f)(x,\xi)|\le C_{\alpha a}
\<\xi\>^{m-\rho|\alpha|+\delta|a|},\ \forall(x,\xi)\in \Xi\Big\}.
\end{equation*}
\end{definition}

It is well known that $S^m_{\rho,\delta}(\Xi)$ is a Fr\'echet space under the
family of semi-norms $\{\sigma^{\alpha a}_m\}_{\alpha,a \in \N^n}$, where
$\sigma^{\alpha a}_m : S^m_{\rho,\delta}(\Xi)\rightarrow\R_+$ is defined by
\begin{equation*}
\sigma^{\alpha a}_m(f)  :=  \underset{(x,\xi)\in\Xi}{\sup} \big\{
\expval{\xi}^{-m+\rho|\alpha|-\delta|a|}|
(\partial^a_x\partial^\alpha_\xi f)(x,\xi)|\big\}.
\end{equation*}

\begin{remark}\label{surlesindices}
The product formula \eqref{sifil} has been proved in
\cite[Thm.~2.2]{IMP1} under the assumption $0\leq \delta<\rho\leq
1$. But the special case $\rho=\delta=0$ is a consequence of the
statement contained in \cite{IMP2}.
\end{remark}

\subsection{Magnetic pseudodifferential operators}\label{mpsd}

Being a closed $2$-form in $\X$, the magnetic field can be written as $B=dA$ for some $1$-form $A$ called
{\it vector potential}.
Any equivalent choice $A'=A+d\psi$, with $\psi:\X\to \R$ of
suitable smoothness, will give the same magnetic field. It is easy to see that
if $B$ is of class $C^\infty_{\rm{pol}}(\X)$, then $A$ can be chosen in the
same class, which is tacitly assumed in the sequel. For example, the
vector potential in the so-called ``transversal gauge'' satisfies this property.

For any vector potential $A$ defining the magnetic field $B$, and for
$x,y \in \X$ let us set
\begin{equation*}
\Gamma^A([x,y]):=\int_{[x,y]}A
\end{equation*}
for the circulation of $A$ along the linear segment $[x,y]$
(integration of a $1$-form through a $1$-simplex).
We can then define for $u: \X \to \C$ the map:
\begin{equation}\label{op}
\big[ \Op^A(f) u \big] (x) :=  \int_\X \de y \int_{\X^*}  \de \eta \,
e^{i(x-y) \cdot \eta} e^{-i \Gamma^A([x,y])}
 f \bigl ( \tfrac{x+y}{2} , \eta \bigr ) u(y).
\end{equation}
For $A=0$ one recognizes the Weyl quantization, associating to functions or
distributions on $\Xi$ linear operators acting on function spaces on $\X$.
Suitably interpreted and by using rather simple duality arguments,
$\Op^A$ defines a representation of the $^*$-algebra $\MB(\Xi)$ by
linear continuous operators $:\S(\X)\rightarrow\S(\X)$. This
means that $\Op^A(f \sh g)=\Op^A(f) \, \Op^A(g)$ and
$\Op^A(\overline{f})=\Op^A(f)^*$
for any $f,g\in\MB(\Xi)$. In addition, $\Op^A$ restricts to an
isomorphism from $\S(\Xi)$ to $\B\big(\S'(\X),\S(\X)\big)$ and extends to an
isomorphism  from $\S'(\Xi)$ to $\B\big(\S(\X),\S'(\X)\big)$, where
$\B(\mathcal R,\mathcal T)$
is the family of all linear and continuous operators between the topological
vector spaces $\mathcal R$ and $\mathcal T$.

An important property of \eqref{op} is {\it gauge covariance}: if $A'=A+d\psi$
defines the same magnetic field as $A$, then $\Op^{A'}(f)=e^{i \psi} \, \Op^A(f)
e^{-i \psi}$. Such a property would not hold for the wrong quantization,
appearing in the literature
\begin{eqnarray*}
\big[ \Op_A(f) u \big](x):=\int_{\X} \de y \int_{\X^*} \de \eta \, e^{i (x -
y) \cdot \eta}  \, f \bigl ( \tfrac{x+y}{2} , \eta-A \bigl (\tfrac{x+y}{2} \bigr ) \bigr ) \, u(y).
\end{eqnarray*}

Another important result is a magnetic version of the Calderon-Vaillancourt theorem:

\begin{theorem}\label{uru}
Assume that the components of the magnetic field belong to
$BC^\infty(\X)$, and let $f \in S^0_{\rho,\rho}(\Xi)$ for some $\rho
\in [0,1)$. Then $\Op^A(f)
\in \mathcal{B} \bigl (L^2(\X) \bigr )$ and we
have the inequality
\begin{equation*}
\bnorm{\Op^A(f)}_{\mathcal{B}(L^2(\X))} \leq c(n) \sup_{\abs{a} \leq
  p(n)} \sup_{\abs{\alpha} \leq p(n)}
\sup_{(x,\xi) \in \Xi} \expval{\xi}^{\rho(|\alpha|-|a|)} \babs{\partial_x^a
\partial_{\xi}^{\alpha} f(x,\xi)},
\end{equation*}
where $c(n)$ and $p(n)$ are constants depending only on the dimension of the configuration space.
\end{theorem}

\subsection{Symbol spaces with coefficients in $\A$}\label{mirabilis}

We first introduce the coefficients $C^*$-algebra $\A$, which can be thought of as a way to encode the
behavior of the magnetic fields and of the configurational part of the symbols.

Let $\A$ be a unital $C^*$-subalgebra of $BC_u(\X)$, the
set of bounded and uniformly continuous functions on $\X$. Depending on the
context, the $L^\infty$-norm of this algebra will be denoted either by
$\|\cdot\|_\A$ or by $\|\cdot\|_\infty$. We shall always assume that $\A$ is
stable by translations, {\it i.e.}~$\theta_x(\varphi):=\varphi(\cdot +x) \in
\A$ for all $\varphi \in \A$ and $x \in \X$, and sometimes we ask that $C_0(\X)$ is contained in
$\A$. Here, $C_0(\X)$ denotes the algebra of continuous functions on $\X$ that
vanish at infinity.

The following definition is general and applies to any $C^*$-algebra $\A$ endowed with an action of $\X$.

\begin{definition}
Let us define $\A^{\infty} := \bigl \{ \varphi \in \A \mid \hbox{the
map }\X\ni x \mapsto {\theta_x}(\varphi) \in \A \ {\rm is} \
C^\infty \bigr \}$. For $a \in \N^n$ we set
\begin{enumerate}[(a)]
\item $\delta^a:\A^\infty \ni \varphi \mapsto  \delta^a(\varphi):=
\partial^a_x\big(\theta_x(\varphi)\big)\big|_{x=0} \in \A^\infty$,
\item $s^a:\A^\infty \ni \varphi \mapsto
  s^a(\varphi):=\|\delta^a(\varphi) \|_\mathcal A \in \R_+$.
\end{enumerate}
\end{definition}

It is known that $\A^\infty$ is a dense $^*$-subalgebra of $\A$, as well as a Fr\'echet $^*$-algebra with the family of
semi-norms $\{s^a \mid a \in \N^n \}$. But our setting is quite special: $\A$
is an abelian $C^*$-algebra composed of bounded and uniformly continuous
complex functions defined on the group $\X$ itself. The easy proof of the next result is left to the reader.

\begin{lemma}\label{achiu}
 $\A^\infty$ coincides with $\bigl \{ \varphi\in
 C^\infty(\X)\mid\partial^a \varphi\in \A, \ \forall a \in \N^n \bigr
 \}$. Furthermore, for any $a \in \N^n$ and $\varphi \in \A^{\infty}$,
 one has $\delta^a(\varphi) = \partial^a_x\varphi$.
\end{lemma}

We now introduce the anisotropic version of the H\"ormander classes of
symbols, {\it cf.}~also \cite{Ba1,Ba2,CMS,Co,Sh1}. For any $f:\Xi\rightarrow \C$ and $(x,\xi)\in\Xi$, we will often write $f(\xi)$ for
$f(\cdot, \xi)$ and $[f(\xi)](x)$ for $f(x,\xi)$. In that situation, $f$ will
be seen as a function on $\X^*$ taking values in some space of functions
defined on $\X$.

\begin{definition}
The space $S^m_{\rho,\delta} (\X^*;\A^\infty)$ of
\emph{$\A$-anisotropic symbols of order $m$ and type $(\rho,\delta)$}
is
\begin{equation*}
\Big\{f\in C^\infty(\X^*;\A^\infty)\mid \forall \alpha,a\in
\N^n,\exists\  C_{\alpha a}<\infty\ \hbox{\rm s.t.}~s^a
[(\partial^\alpha_\xi f)(\xi)]\le C_{\alpha a}
\<\xi\>^{m-\rho|\alpha|+\delta|a|},\ \forall\xi\in \X^*\Big\}.
\end{equation*}
\end{definition}

Due to the very specific nature of the $C^*$-algebra $\A$, we have again some
simplifications:

\begin{lemma}\label{rachiu}
The following equality holds:
\begin{equation}\label{symbol}
S^m_{\rho,\delta}(\X^*;\A^\infty) = \bigl \{ f \in
S^m_{\rho,\delta}(\Xi) \mid  (\partial^a_x\partial^\alpha_\xi f)(\xi)
\in \A , \ \forall \xi\in\X^* \hbox{ \rm and } \alpha,a\in\mathbb N^n
\bigr \}.
\end{equation}
\end{lemma}

\begin{proof}
First we notice that the conditions
\begin{equation*}
s^a[(\partial^\alpha_\xi f)(\xi)]\le C_{\alpha a}
\<\xi\>^{m-\rho|\alpha|+\delta|a|}, \ \forall \xi \in \X^*
\quad \hbox{ and } \quad
\left|(\partial_x^a\partial^\alpha_\xi f)(x,\xi)\right|\le C_{\alpha
a} \<\xi\>^{m-\rho|\alpha|+\delta|a|}, \
\forall (x,\xi)\in \Xi
\end{equation*}
are identical. On the other hand, by Lemma \ref{achiu},
\begin{equation*}
(\partial^\alpha_\xi f)(\xi)\in\A^\infty\ \Longleftrightarrow\
(\partial_x^a\partial^\alpha_\xi f)(\xi)\in\A,\ \
\forall a\in\mathbb N^n.
\end{equation*}
It thus follows that $S^m_{\rho,\delta} (\X^*;\A^\infty)$ is included in the
r.h.s.~of \eqref{symbol}, and we are then left with proving that if $f\in
S^m_{\rho,\delta}(\Xi)$ and $(\partial^\alpha_\xi f)(\xi)\in\A^\infty$ for all
$\alpha$ and $\xi$, then $f\in C^\infty (\T)$.

We first show that $f:\X^*\rightarrow \A^\infty$ is differentiable, that is for
each $a \in \N^n$:
\begin{equation*}
s^a\left[\frac{1}{t}\left[f(\xi+te_j)-f(\xi)\right]-(\partial_{\xi_j}f)(\xi)\right]
\underset{t\rightarrow0}{\longrightarrow} 0,\ \ \forall j=1,\dots,n,
\end{equation*}
where $e_1,\dots,e_n$ is the canonical base in $\X^*\cong \R^n$. Indeed, we
have for $t>0$:
\begin{eqnarray*}
&& \underset{x\in \X}{\sup}\left|\frac{1}{t}[(\partial^a_x
  f)(x,\xi+te_j)-(\partial^a_x f)
(x,\xi)]-(\partial^a_x\partial_{\xi_j} f)(x,\xi)\right| \\
= && \underset{x\in\X}{\sup}\left|\frac{1}{t}\int^t_0 \de s\int^s_0
  \de u(\partial^a_x\partial^2_{\xi_j}f)(x,\xi+ue_j)
\right| \\
\le && \underset{x\in\X}{\sup}\frac{1}{t}\int^t_0 \de s\int^s_0 \de
u\,C_a\<\xi+ue_j\>^{m-2\rho+\delta|a|} \\
\le && C'_a\<\xi\>^{m-2\rho+\delta|a|}\frac{1}{t}\int^t_0 \de s
\int^s_0 \de u \,\<u\>^{|m-2\rho+\delta|a||} \\
\le && C''_a\<\xi\>^{m-2\rho+\delta|a|}\frac{1}{t}(t^2-0)
\quad\underset{t\rightarrow 0}{\longrightarrow} 0,
\end{eqnarray*}
and similarly for $t<0$. We can continue to apply this procedure to the
resulting derivative $\partial_{\xi_j} f\in S^{m-\rho}_{\rho,\delta}(\Xi)$ and
finish the proof by recurrence.
\end{proof}

In particular, for $\A=BC_u(\X)$, it is easy to see that
\begin{eqnarray*}
BC_u(\X)^\infty &=& \bigl \{\varphi\in
C^\infty(\X)\mid \partial^a \varphi\in BC_u(\X), \ \forall a \in\N^n
\bigr \} \\
&=& \bigl \{ \varphi\in C^\infty(\X)\mid \partial^a \varphi\in BC(\X),
\ \forall a \in \N^n \bigr \}=:BC^\infty(\X).
\end{eqnarray*}

Then it follows from the previous lemma that
\begin{equation*}
    S^m_{\rho,\delta} \bigl ( \X^*;BC_u(\X)^{\infty} \bigr
    )=S^m_{\rho,\delta}\big(\X^*;BC^\infty(\X)\big)
    =S^m_{\rho,\delta}(\Xi).
\end{equation*}

\begin{proposition}\label{ionel}
\begin{enumerate}[(a)]
\item $S^{m}_{\rho,\delta}(\T)$ is a closed subspace of the Fr\'echet
  space $S^{m}_{\rho,\delta}(\Xi)$.
\item For any $m_1, m_2 \in \R$, $S^{m_1}_{\rho,\delta}(\T)\cdot
  S^{m_2}_{\rho,\delta}(\T)
\subset S^{m_1+m_2}_{\rho,\delta}(\T)$,
\item For any $\alpha,a \in \N^n$, $\partial^a_x\partial^\alpha_\xi
  S^{m}_{\rho,\delta}(\T)
\subset S^{m-\rho |\alpha|+\delta |a|}_{\rho,\delta}(\T)$.
\end{enumerate}
\end{proposition}

\begin{proof}
(a) We have to show that if $f_n\in S^{m}_{\rho,\delta}(\T)$, $f\in
S^{m}_{\rho,\delta}(\Xi)\,$ and $\sigma^{\alpha a}_m(f_n-f)
\underset{n\rightarrow\infty}{\longrightarrow}0\,$, then
$(\partial_x^a\partial_\xi^\alpha f)(\xi)\in\A$ for all $\alpha,a, \xi$. But
since $\A$ is closed, it is enough to show that for any $a,\alpha\in\N^n$, the
following statement holds: if $g_n \in S^{m}_{\rho,\delta}(\Xi)$ and
$\sigma^{\alpha a}_m(g_n) \underset{n\rightarrow\infty}{\longrightarrow}0$,
then $\|(\partial_x^a\partial_\xi^\alpha g_n) (\xi)\|_\infty
\underset{n\rightarrow\infty}{\longrightarrow}0,\ \ \forall \xi\in\X^*$. This
follows from the definition of $\sigma^{\alpha a}_m$.

Statement (b) follows by applying Lemma \ref{rachiu}, Leibnitz's rule and the
fact that $\A$ is an algebra. Statement (c) is a direct consequence of Lemma
\ref{rachiu}.
\end{proof}

\subsection{Symbol composition}\label{position}

In this section we study the product of two symbols by the
composition law $\sh$ defined in \eqref{produitMoyal}.
For simplicity, we introduce $\ob$ and $\gb$ (low indices) by the relations
\begin{equation*}
\ob(x,y,z)=e^{-i\gb(x,y,z)}:=e^{-i\Gamma^B(\<x-y-z,x+y-z,x-y+z\>)}.
\end{equation*}
One has explicitly
\begin{equation}\label{explicitformula}
\gb(x,y,z)  = \sum_{j,k=1}^n y_j\;\!z_k
\int_0^2 \de s \int_0^1 \de t\;\!s\;\!B_{jk}\big(x+(s-st-1)\;\!y
+(st-1)\;\!z \big)
\end{equation}
and \eqref{produitMoyal} reads
\begin{equation}\label{Moyal2}
[f\sh g](X):=4^n\int_\Xi \de Y \int_\Xi
\de Z \;\!e^{-2i\sigma(Y,Z)}\;\!\ob(x,y,z)\;\!f(X-Y)\;\!g(X-Z).
\end{equation}

We state the main result of this section :

\begin{theorem}\label{babana}
Assume that the each component $B_{jk}$ belongs to $\A^\infty$. Then, for any
$m_1,m_2\in \R$ and $0\leq \delta<\rho\leq 1$ or $\rho=\delta=0$,
one has
\begin{equation}\label{eqmain}
S^{m_1}_{\rho,\delta}(\T)\;\sh\; S^{m_2}_{\rho,\delta}(\T)\;\subset\;
S^{m_1+m_2}_{\rho,\delta}(\T).
\end{equation}
\end{theorem}

Before proving this theorem, we need a technical lemma.

\begin{lemma}\label{sticloasa}
Assume that the each component $B_{jk}$ belongs to $\A^\infty$. Then, for all
$a,b,c\in\N^n$ and all $x,y,z \in \X$, one has:
\begin{enumerate}
\item[(a)] $\big(\partial_x^a\;\! \partial_y^b\;\! \partial_z^c\;\!
\gb\big) (\cdot,y,z)\in \A$,
\item[(b)] $\big(\partial_x^a\;\! \partial_y^b\;\! \partial_z^c\;\!
\ob\big) (\cdot,y,z)\in \A$,
\item[(c)] $\ \big|(\partial_x^a\;\! \partial_y^b\;\! \partial_z^c\;\!
\ob) (x,y,z)\big| \le C_{abc}\big(\<y\>+
\<z\>\big)^{|a|+|b|+|c|}$.
\end{enumerate}
\end{lemma}

\begin{proof}
The expressions $\big(\partial_x^a\;\! \partial_y^b\;\!
\partial_z^c\;\! \gb\big) (\cdot,y,z)$ can be explicitly
calculated by using \eqref{explicitformula}, (a) follows from the completeness of
$\A$ and (b) easily follows from (a). Statement (c) is borrowed from \cite{IMP1}.
\end{proof}

\begin{proof}[Proof of Theorem \ref{babana}]
Since the components of the magnetic field belong to
$BC^\infty(\X)\subset C^\infty_{\hbox{\tiny \rm pol}}
(\X)$, it follows from Lemma \ref{rachiu} and \cite[Lem.~1.2]{IMP1}
that $S^{m_j}_{\rho,\delta}(\T)\subset
S^{m_j}_{\rho,\delta}(\Xi) \subset \MB(\Xi)$, for $j \in \{1,2\}$,
and thus the $\sh$-product in \eqref{eqmain} is
well defined in $\MB(\Xi)$, as explained in Section
\ref{veil}. Under the additional hypothesis that
$B_{jk}\in BC^\infty(\X)$, it has even been proved in
\cite[Thm.~2.2]{IMP1} (see also Remark \ref{surlesindices}) that the
product belongs to $S^{m_1+m_2}_{\rho,\delta}(\Xi)$  and can also be
defined by the usual oscillatory integral techniques.
Thus, thanks to Lemma \ref{rachiu}, it only remains to show that for
any $\alpha, a \in \N^n$, $f\in S^{m_1}_{\rho,\delta}(\T)$ and $g\in S^{m_2}_{\rho,\delta}(\T)$, the expression
$\big[\partial_x^a\partial _\xi^\alpha(f\sh g)\big] (\xi)$ belongs to $\A$, for all $\xi \in \X^*$.

For that purpose, let $\alpha^1,\alpha^2,a^0,a^1,a^2 \in \N^n$ with
$\alpha^1 + \alpha^2=\alpha$ and $a^0+a^1+a^2=a$.
We define $F_{\alpha^1 a^1}:=\partial_x^{a^1}\partial_\xi^{\alpha^1}f
\in S^{p_1}_{\rho,\delta}(\T)$,
$G_{\alpha^2 a^2}:=\partial_x^{a^2}\partial_\xi^{\alpha^2}g\in
S^{p_2}_{\rho,\delta}(\T)$ and
$\Ob^{a^0}:=\partial_x^{a^0}\ob$. Then
$p_j=m_j-\rho|\alpha^j|+\delta|a^j|$ for $j \in \{1,2\}$ and
$\Ob^{a^0}$ satisfies the properties of Lemma \ref{sticloasa}. We have
to study the $x$-behavior of the expression
\begin{eqnarray}\label{eqhoribilis}
\nonumber \big[\partial_x^a\partial_\xi^\alpha(f\sh g)\big](x,\xi)
=\underset{\underset{\alpha^1+\alpha^2=\alpha}
{a^0+a^1+a^2=a}}{\sum}
&& C_{a^0a^1a^2}^{\alpha^1\alpha^2}\int_{\X}\de y \int_{\X}\de z
\int_{\X^*}\de \eta\int_{\X^*} \de \zeta\,
e^{-2i z \cdot \eta} \, e^{2iy \cdot \zeta}\, \Ob^{a^0}(x,y,z)\ \cdot \\
&& \cdot \ F_{\alpha^1 a^1}(x-y,\xi-\eta)\,G_{\alpha^2 a^2}(x-z,\xi-\zeta)\ .
\end{eqnarray}
The precise definition of these integrals involves rewriting the
factors $e^{-2iz \cdot \eta} \, e^{2iy\cdot \zeta}$ as
\begin{equation}\label{froufrou}
\<y\>^{-2q}\<z\>^{-2q}\<D_\zeta\>^{2q}\<D_\eta\>^{2q}
\<\eta\>^{-2p}\<\zeta\>^{-2p}\<D_z\>^{2p}\<D_y\>^{2p}\big(e^{-2iz\cdot \eta} \, e^{2iy \cdot \zeta}\big)
\end{equation}
where $D:=\frac{1}{2i}\partial$ and $p,q\in \N$, and integrating by parts.
So the r.h.s.~of \eqref{eqhoribilis} contains the integrals
\begin{eqnarray*}
&\int_{\X}\de y\int_{\X}\de z\int_{\X^*}\de \eta \int_{\X^*}\de \zeta \,
e^{-2iz \cdot \eta} \, e^{2iy \cdot \zeta}\,
\<\eta\>^{-2p}\<\zeta\>^{-2p}\ \cdot & \\
& \cdot\ \<D_z\>^{2p}\<D_y\>^{2p}\Big\{\<y\>^{-2q}\<z\>^{-2q} \,
\Ob^{a^0}(x,y,z)\<D_\zeta\>^{2q}\<D_\eta\>^{2q}
\Big[F_{\alpha^1 a^1}(x-y,\xi-\eta)\,G_{\alpha^2
a^2}(x-z,\xi-\zeta)\Big]\Big\}, &
\end{eqnarray*}
which will be proved now to be absolutely convergent for $p,q$ large
enough.

For this, one has to estimate
\begin{eqnarray}\label{hororo}
\nonumber & \<\eta\>^{-2p}\<\zeta\>^{-2p}\,
\<D_z\>^{2p}\<D_y\>^{2p}\Big\{\<y\>^{-2q}\<z\>^{-2q}\,
\Ob^{a^0}(x,y,z)\<D_\zeta\>^{2q}\<D_\eta\>^{2q} \Big[F_{\alpha^1
a^1}(x-y,\xi-\eta)\,G_{\alpha^2 a^2}(x-z,\xi-\zeta)\Big]\Big\} &
\\ & =\<\eta\>^{-2p} \<\zeta\>^{-2p} \<z\>^{-2q} \<y\>^{-2 q}
\underset{|\beta^1|\leq q, \,|\beta^2|\leq q}
{\underset{|c^1|+|c^2|+|c^3|=2p}{\underset{|b^1|
+|b^2|+|b^3|=2p}{\sum}}}C^{c^1 c^2 c^3 \beta^2}_{ b^1 b^2
b^3\beta^1}\,\varphi_{q c^1}(z)\, \psi_{q b^1}(y)\ \cdot & \\ &
\nonumber \cdot\ \big(\partial^{b^2}_{y}
\partial^{c^2}_z\Ob^{a^0}\big) (x,y,z)\, \big(\partial^{b^3}_y
\partial^{2\beta^1}_{\xi}F_{\alpha^1 a^1}\big)
(x-y,\xi-\eta)\,
\big(\partial^{c^3}_z\partial^{2\beta^2}_{\xi}G_{\alpha^2 a^2}\big)
(x-z,\xi-\zeta), &
\end{eqnarray}
where $b^1,b^2,b^3,c^1,c^2,c^3,\beta^1,\beta^2 \in \N^n$, and
$\varphi_{q c^1}$ and $\psi_{q b^1}$ are bounded
functions produced by derivating the factors $\<z\>^{-2q}$ and
$\<y\>^{-2q}$, respectively.
By using the estimates obtained in Lemma \ref{sticloasa} for
$\Ob^{a^0}$, and the a priori estimates on
$F_{\alpha^1 a^1}$ and $G_{\alpha^2 a^2}$, the absolute value of \eqref{hororo} is dominated by
\begin{eqnarray*}
& C_{pq}\,\<\eta\>^{-2p}\<\zeta\>^{-2p}\<z\>^{-2q}\<y\>^{-2q}
\underset{|\beta^1|\leq q,\,|\beta^2|\leq q}
{\underset{|c^1|+|c^2|+|c^3|=2p}{\underset{|b^1|+|b^2|+|b^3|=2p}{\sum}}}
\big(\<y\>+\<z\>\big)^{|a^0|+|b^2|+|c^2|}
\<\xi-\eta\>^{p_1-2\rho |\beta^1|+\delta|b^3|}
\<\xi-\zeta\>^{p_2-2\rho |\beta^2|+\delta|c^3|} &\\
& \leq C_{pq}(\xi)\,\<\eta\>^{-2p(1-\delta)+p_1}\,
\<\zeta\>^{-2p(1-\delta)+p_2}
\,\<y\>^{-2q+|a|+4p}\,\<z\>^{-2q+|a|+4p}. &
\end{eqnarray*}
Since $(1-\delta)>0$, the factors involving $\eta$ and $\zeta$ will be
integrable for $p$ large enough. Fixing a
suitable $p$, for an even larger $q$ we also ensure integrability in $y$ and $z$.

To sum up, $\big[\partial_x^a\partial_\xi^\alpha(f\sh g)\big](x,\xi)$
is given by an absolutely convergent integral,
the integrand being a function of $x$ which belongs to $\A$ for all
values of $\xi,y,\eta,z,\zeta$. It is easy to conclude,
by the Dominated Convergence Theorem, that the map $x\mapsto
\big[\partial_x^a\partial_\xi^\alpha(f\sh g)\big](x,\xi)$
also belongs to $\A$, and this finishes the proof.
\end{proof}

\subsection{Asymptotic developments}\label{simptotic}

In this section we simplify and generalize to $\A$-valued symbols the asymptotic
expansion of the magnetic product of two symbols already derived in
\cite{IMP1}. We refer to \cite{Le} for parameter-dependent developments.

For any multi-index $\alpha \in \N^m$, we use the
notation $\alpha!=\alpha_1! \dots \alpha_m!$. For shortness we shall
also write $\a:=(a,\alpha)$ and $\b:=(b,\beta)$, with $\a,\b \in \N^{2n}$.

\begin{theorem}\label{bibigul}
Assume that the each component $B_{jk}$ belongs to $\A^\infty$ and
let $m_1,m_2 \in \R$ and $\rho \in (0,1]$. Then for any $f\in
S^{m_1}_{\rho,0}(\T)$, $g\in S^{m_2}_{\rho,0}(\T)$ and $N \in \N^*$ one has
\begin{equation*}
f\sh g =\sum_{l=0}^{N-1}h_l + R_N,
\end{equation*}
with
\begin{equation*}
h_l = \underset{|\alpha|+|\beta|=l}{\underset{a\le\beta,
b\le\alpha}{\underset{a,b,\alpha,\beta\in\mathbb N^n}
{\sum}}}h_{\a,\b}\in S^{m_1+m_2-\rho l}_{\rho,0}(\T)
\end{equation*}
and
\begin{equation*}
h_{\a,\b} (x,\xi)=C_{\a\b}\big[(\partial^{\beta-a}_y\partial^{
\alpha-b}_z\ob)(x,0,0)\big]\,\!\big[(\partial^a_x\partial^\alpha_\xi
f)(x,\xi)\big]\,\!\big[(\partial^b_x\partial^\beta_\xi g)(x,\xi)\big],
\end{equation*}
and the constants are given by
\begin{equation*}
C_{\a\b}=\left(\frac{i}{2}\right)^l\frac{(-1)^{|a|+|b|+
|\beta|}}{a!b!(\alpha-b)!(\beta -a)!}.
\end{equation*}
The remainder term $R_N$ belongs to $S^{m_1+m_2-\rho N}_{\rho,0}(\T)$.
\end{theorem}

\begin{remark}
If $B=0$, which implies that $\ob =1$, one has
$h_{\a,\b}\ne 0$ only if $a=\beta$ and $b=\alpha$; by setting
$\hat{\a}$ for $(\alpha,a)$, one has $ h_{\a, \hat{\a}}=
\frac{(-1)^{|\alpha|}}
{\a !}\left(\frac{i}{2}\right)^{|\a|}\partial^{\a}f \,\partial^{\hat{\a}}g$.
\end{remark}

Before proving the theorem, we list the first two terms in the development:
\begin{eqnarray*}
\ h_0 &=&\,f g, \\
h_1&=&\frac{i}{2}\{ f,g\}=\frac{i}{2}\sum^n_{j=1}(\partial_{x_j} f\,\partial_{\xi_j} g-\partial_{\xi_j}
f\,\partial_{x_j}g).
\end{eqnarray*}

\begin{proof}[Proof of Theorem \ref{bibigul}.]
In the formula \eqref{Moyal2} we shall use the Taylor series
\begin{equation*}
(f\otimes g)(X-Y,X-Z)=\underset{|(\a,\b)|<N}{\sum}
\frac{(-1)^{|(\a,\b)|}}{(\a,\b)!}(Y,Z)^{(\a,\b)}
[\partial^{(\a,\b)}(f\otimes g)](X,X)+ r_{f,g}(X,Y,Z),
\end{equation*}
where the remainder $r_{f,g}$ will be specified later.
It follows that
\begin{equation*}
f\sh g = \underset{|(\a,\b)|<N}{\sum}h_{\a,\b}+R_N,
\end{equation*}
with
\begin{equation*}
h_{\a,\b}(X)=\frac{(-1)^{|(\a,\b)|}}{(\a,\b)!}
[\partial^{(\a,\b)}(f\otimes g)](X,X)\;4^n\int_\Xi \de Y \int_\Xi
\de Z\,(Y,Z)^{(\a,\b)}e^{-2i\sigma( Y,Z)}\,\ob (x,y,z).
\end{equation*}
In other words, one has
\begin{equation*}
h_{\a,\b}=\frac{(-1)^{|a|+|b|+|\alpha|+|\beta|}}{a!b!\alpha
  !\beta !} [\partial^a_x \partial^\alpha_\xi f]\,
[\partial^b_x\partial^\beta_\xi g]\,\Omega_{\a,\b},
\end{equation*}
with $\Omega_{\a,\b}(x)$ given by
\begin{eqnarray*}
&& 4^n\int_\X \de y \int_\X \de z\,y^a z^b\ob (x,y,z)
\Big[\int_{\X^*} \de \eta\, e^{-2i z\cdot \eta}
\eta^\alpha\Big]\Big[ \int_{\X^*} \de\zeta\,
e^{2i y\cdot\zeta}\zeta^\beta\Big] \\
&=& \frac{(-i)^{\abs{\alpha}}
i^{\abs{\beta}}}{2^{\abs{\alpha}+\abs{\beta}}} \partial^{\beta}_y
\partial^{\alpha}_z  \bigl \{ y^a z^b \ob (x,y,z) \bigr \} \vert_{y=z=0}.
\end{eqnarray*}
The following factor vanishes unless $b\le\alpha$ and $a\le\beta$:
$$
\partial^\beta_y\partial^\alpha_z\left\{y^a z^b\ob (x,y,z)\right\}\mid_{y=z=0}
=\frac{\alpha!\beta!}{(\alpha-b)!(\beta-a)!}(\partial^{\beta-a}_y
\partial^{\alpha-b}_z\ob) (x,0,0)\,.
$$
So, restricting to the case $b\le \alpha$ and $a\le\beta$, we can write:
\begin{equation*}
h_{\a,\b}(x,\xi)=\frac{(-1)^{|a|+|b|}\,i^{|\alpha|} (-i)^{|\beta|}}{a!b!(\alpha-b)!(\beta-a)!}
\left(\frac{1}{2}\right)^{|\alpha|+|\beta|}\big[(\partial^{\beta-a}_y\partial^{
\alpha-b}_z\ob)(x,0,0)\big]\,\!\big[(\partial^a_x\partial^\alpha_\xi
f)(x,\xi)\big]\,\!\big[(\partial^b_x\partial^\beta_\xi g)(x,\xi)\big].
\end{equation*}
By Proposition \ref{ionel} and Lemma \ref{sticloasa},  one finally obtains that
$h_{\a,\b}\in S^{m_1+m_2-\rho(|\alpha|+|\beta|)}_{\rho,0}(\T)$.

We now treat the remainder $R_N(X)$ given by
\begin{eqnarray*}
&4^n \int_\Xi \de Y \int_\Xi \de Z e^{-2i\sigma( Y,Z)}\ob(x,y,z)
\underset{|(\a,\b)|=N}{\sum}\frac{(Y,Z)^{(\a,\b)}}{(\a,\b)!}
N\int^1_0 \de \tau (1-\tau)^{N-1}[\partial^{(\a,\b)}(f\otimes g)](X-\tau Y,X-\tau Z)& \\
&=\underset{|a|+|b|+|\alpha|+|\beta|=N}{\sum}\frac{4^n
N}{a!b!\alpha!\beta!}\int^1_0  \de \tau (1-\tau)^{N-1}
\int_{\X}\de y \int_{\X}\de z \int_{\X^*}\de \eta \int_{\X^*}\de \zeta
\, \ob(x,y,z)\cdot & \\
&\cdot \; y^a z^b \eta^\alpha \zeta^\beta
e^{-2i\sigma(Y,Z)}\;[\partial^a_x \partial^\alpha_\xi f](x-\tau
y,\,\xi -\tau\eta) \;[\partial^b_x\partial^\beta_\xi g](x-\tau z,\,\xi - \tau z). &
\end{eqnarray*}
In order to show that this term belongs to $S^{m_1+m_2-\rho
N}_{\rho,0}(\T)$, we take into account
\begin{equation*}
y^a z^b \eta^\alpha \zeta^\beta e^{-2i\sigma( Y,Z)}=\frac{1}{(2i)^{|a|} (-2i)^{|\alpha|} (-2i)^{|b|}
(2i)^{|\beta|}}\;\!\partial^a_\zeta \;\!\partial^\alpha_z\;\!
\partial^b_\eta\;\! \partial^\beta_y\;\!  e^{-2i\sigma(Y,Z)},
\end{equation*}
and introduce it into $R_N (X)$ which can then be rewritten as
\begin{equation*}
\underset{|a|+|b|+|\alpha|+|\beta|=N}{\sum}\frac{4^n N (-1)^{|a|+|\beta|}}{a!b!\alpha!\beta!(2i)^N}
\int^1_0 \de \tau(1-\tau)^{N-1} \int_\X \de y \int_\X \de z \int_{\X^*} \de \eta \int_{\X^*} \de \zeta \,e^{-2i\sigma(Y,Z)}
\phi^{\tau}_{\a,\b}
(X,Y,Z),
\end{equation*}
with
\begin{eqnarray*}
\phi^{\tau}_{\a,\b}(X,Y,Z)&:=&\partial^a_\zeta\partial^\alpha_z\partial^b_\eta
\partial^\beta_y \left[\ob (x,y,z) [\partial^a_x\partial^\alpha_\xi
f](x-\tau y,\xi-\tau\eta)  \;[\partial^b_x\partial^\beta_\xi g]
(x-\tau z, \xi-\tau\zeta) \right]\\
&=&\,\underset{\alpha'\le\alpha}{\sum}\,\underset{\beta'\le\beta}{\sum}
\noverk{\alpha}{\alpha'} \noverk{\beta}{\beta'}
[\partial^{\alpha-\alpha'}_z\partial^{\beta-\beta'}_y\ob](x,y,z)\;
\partial^{\beta'}_y\partial^b_\eta\big[
\big(\partial^a_x\partial^\alpha_\xi f\big)(x-\tau y,\xi-\tau\eta)\big] \ \cdot \\
&& \cdot\  \partial^{\alpha'}_z\partial^a_\zeta
\big[\big(\partial^b_x \partial^\beta_\xi g\big) (x-\tau z,\xi-\tau\zeta)\big] \\
&=&\underset{\alpha'\le\alpha}{\sum}\,\underset{\beta'\le\beta}{\sum}
\noverk{\alpha}{\alpha'} \noverk{\beta}{\beta'}
(-\tau)^{|b|+|a|+|\beta'|+|\alpha'|}\;[\partial^{\alpha-\alpha'}_z\partial^{\beta-\beta'}_y\ob]
(x,y,z)\ \cdot \\
&&\cdot\ [\partial^{a+\beta'}_x\partial^{\alpha+b}_\xi f](x-\tau
y,\xi-\tau\eta)\;  [\partial^{b+\alpha'}_x
\partial^{a+\beta}_\xi g] (x-\tau z,\xi-\tau\zeta).
\end{eqnarray*}
So we have
\begin{equation}\label{eqR}
R_N(X)=\underset{|a|+|b|+|\alpha|+|\beta|=N}{\underset{\alpha'\le\alpha,\,
\beta'\le\beta}{\underset{a,b,\alpha,
\beta,\alpha',\beta'}{\sum}}}\int^1_0 \de \tau\,{\rm
pol}_{\a,\b}^{\alpha',\beta'} (\tau)\  I^{\alpha',\beta'}_{\tau,\a,\b}(X),
\end{equation}
where ${\rm pol}_{\a,\b}^{\alpha',\beta'} :[0,1]\to\C$ are polynomials and
\begin{eqnarray*}\label{iii}
I^{\alpha',\beta'}_{\tau,\a,\b}(X):=&& \int_\X \de y \int_\X \de z
\int_{\X^*} \de \eta \int_{\X^*} \de \zeta \,e^{-2i\sigma(Y,Z)}\;
[\partial^{\alpha-\alpha'}_z\partial^{\beta-\beta'}_y\ob]
(x,y,z)\ \cdot \\
&&\cdot\ [\partial^{a+\beta'}_x\partial^{\alpha+b}_\xi f](x-\tau
y,\xi-\tau\eta)\; [\partial^{b+\alpha'}_x
\partial^{a+\beta}_\xi g] (x-\tau z,\xi-\tau\zeta).
\end{eqnarray*}
Retaining only its essential features, we shall rewrite this last
expression as
\begin{equation*}
I_\tau(X):=\int_\X \de y \int_\X \de z \int_{\X^*} \de \eta
\int_{\X^*} \de \zeta  \,e^{-2i\sigma(Y,Z)}\;\sb(x,y,z)\;F(x-\tau
y,\xi-\tau\eta) \;G(x-\tau z,\xi-\tau\zeta)\ .
\end{equation*}

In order to show that $R_N$ belongs to $S^{m_1+m_2 - \rho N}_{\rho,0}(\Xi)$, let us
calculate
$\partial^{d}_{x}\partial^{\delta}_{\xi}I_{\tau}$. Actually, by
using \eqref{froufrou}, the oscillatory integral
definition of $\partial_x^d\partial_\xi^\delta I_\tau$ is
\begin{equation*}
[\partial^{d}_{x}\partial^{\delta}_{\xi}I_{\tau}](X)
=\,\underset{\delta^1+\delta^2=\delta}{\underset{d^0+d^1+d^2=d}{\sum}}
C^{\delta^1 \delta^2}_{d^0 d^1 d^2} \int_\X \de y \int_\X \de z
\int_{\X^*} \de \eta \int_{\X^*} \de \zeta\; e^{-2i\sigma(Y,Z)}
\;L^{\tau, \delta^1, \delta^2}_{p, q, d^0, d^1, d^2}(X,Y,Z)\ ,
\end{equation*}
where, for suitable integers $p,q$, the expression $L^{\tau, \delta^1,
\delta^2}_{p, q, d^0, d^1, d^2}(X,Y,Z)$ is given by
\begin{eqnarray*}
&&\<\eta\>^{-2p}\<\zeta\>^{-2p}\<D_y\>^{2p}\<D_z\>^{2p}\Big[
\<y\>^{-2q}\<z\>^{-2q} [\partial^{d^0}_{x}\sb](x,y,z)\ \cdot \\
&& \cdot \ \<D_\eta\>^{2q}\<D_\zeta\>^{2q}\;[\partial^{d^1}_x\partial^{\delta^1}_\xi
F](x-\tau y,\xi-\tau\eta)\;[\partial^{d^2}_x\partial^{\delta^2}_\xi G](x-\tau z,\xi-\tau\zeta)\Big] \\
& = & \<\eta\>^{-2p}\<\zeta\>^{-2p}\<y\>^{-2q}\<z\>^{-2q}\underset{|q^1|\leq
q,\,|q^2|\leq q}{\underset{|c^1|+|c^2|+|c^3|=2p}{\underset{|b^1|+|b^2|+ |b^3|=2p}{\sum}}}
C^{q^1 q^2 c^1 c^2 c^3}_{b^1 b^2 b^3} \varphi_{q c^1}(z)\;\! \psi_{q
b^1}(y)\;\!  [\partial^{d^0}_x\partial^{b^2}_y\partial^{c^2}_z\sb](x,y,z)\ \cdot \\
&& \cdot(-\tau)^{2|q^1|+2|q^2|+|b^3|+|c^3|}
[\partial^{d^1+b^3}_x\partial^{\delta^1+2q^1}_\xi F]
(x-\tau y,\xi-\tau\eta) \;
[\partial^{d^2+c^3}_x\partial^{\delta^2+2q^2}_\xi G](x-\tau z,\xi-\tau\zeta),
\end{eqnarray*}
where $\varphi_{q c^1}$ and $\psi_{q b^1}$ are bounded
functions produced by derivating the factors $\<z\>^{-2q}$  and
$\<y\>^{-2q}$, respectively. By taking the explicit form of
$\sb,F,G$ and Lemma \ref{sticloasa} into account, one has
\begin{eqnarray*}
\big|L^{\tau, \delta^1, \delta^2}_{p, q, d^0, d^1, d^2}(X,Y,Z)\big|&\leq& C^{\delta^1 \delta^2}_{pq
d^0 d^1 d^2}\; \<\eta\>^{-2p}\<\zeta\>^{-2p}\<y\>^{-2q}\<z\>^{-2q}\
\big(\<y\>+\<z\>\big)^{|d|+|\alpha|+|\beta|+4p}\ \cdot \\
&&\cdot \ \<\xi-\tau\eta\>^{m_1-\rho(|\alpha|+|b|+|\delta^1|+2|q^1|)}
\;\!\<\xi-\tau\zeta\>^{m_2-\rho(|a|+|\beta|+|\delta^2|+2|q^2|)} \\
&\leq& D^{\delta^1\delta^2}_{pq d^0 d^1 d^2}\;\!
\<y\>^{-2q+N+4p+|d|}\;\!\<z\>^{-2q+N+4p+|d|} \;\! \<\xi\>^{m_1+m_2-
\rho (N+|\delta|)}\ \cdot \\
&& \cdot\ \<\eta\>^{-2p+|m_1-\rho
  (|\alpha|+|b|+|\delta^1|)|}\;\!\<\zeta\>^{-2p+ |m_2-\rho (|a|+|\beta|+|\delta^2|)|}.
\end{eqnarray*}
Then, it only remains to insert this estimate into the expression of
$R_N$ given in \eqref{eqR}, and to
observe that by choosing $p$ large enough, one gets absolute
integrability in $\eta$ and $\zeta$. A subsequent choice of $q$
also ensures integrability in $y$ and $z$. The behavior in $\xi$ is finally
the one expected for $\partial^{d}_{x}\partial^{\delta}_{\xi}R_N$.

Thus, we have obtained so far that $R_N$ belongs to $S^{m_1+m_2-\rho
N}_{\rho,0}(\Xi)$. By taking then Theorem \ref{babana} and the properties
of $h_l$ into account, one has
\begin{equation*}
[\partial^{d}_{x}\partial^{\delta}_{\xi}R_N](\cdot,\xi)
= \partial^{d}_{x}\partial^{\delta}_{\xi}\Big[f\sh g
-\sum_{l=0}^{N-1}h_l\Big](\cdot,\xi) \ \in \A
\end{equation*}
for any $\xi \in \X^*$. It finally follows from Lemma \ref{rachiu} that $R_N$
belongs to $S^{m_1+m_2-\rho N}_{\rho,0}(\T)$.
\end{proof}

\section{$C^*$-algebras}\label{brasagain}

\subsection{$C^*$-algebras generated by symbols}\label{acrp1}

We continue to assume that all
components of the magnetic field belong to $\A^\infty$ and let $\H:=L^2(\X)$. 
As already mentioned, we choose a vector potential $A$
that belongs to $C^\infty_{\hbox{\tiny \rm pol}}(\X)$ and thus the map
$\Op^A$ extends to a linear topological isomorphism $\S'(\Xi)\to
\B\big(\S(\X),\S'(\X)\big)$. Since $\B(\H)$ is
continuously imbedded in $\B\big(\S(\X),\S'(\X)\big)$, one can define
\begin{equation*}
\AB:=\big(\Op^A\big)^{-1}[\B(\H)].
\end{equation*}
It is obviously a vector subspace of $\S'(\Xi)$ which only depends on the
magnetic field (by gauge covariance). On convenient subsets, for example on
$\AB\cap \MB(\Xi)$, the transported product from $\B(\H)$ coincides with
$\sh$, and the adjoint in $\B(\H)$ corresponds to the involution
${}^{\sh}$. Endowed with the transported norm $\| f\|_B\,\equiv\,\| f\|_{\AB}
:=\| \Op^A(f)\|_{\B(\H)}$, $\AB$ is a $C^*$-algebra.

With these notations and due to the inclusion $S^m_{\rho,\delta}(\Xi)
\subset S^m_{\delta,\delta}(\Xi)$ for $\delta<\rho$, Theorem \ref{uru}
can be rephrased:

\begin{proposition}\label{vaiancuru}
For any $0\le\delta\le\rho\le 1$ with $\delta\ne 1$, the following continuous
embedding holds:
\begin{equation*}
S^0_{\rho,\delta}(\Xi)\hookrightarrow \AB.
\end{equation*}
\end{proposition}

We shall now define two $\A$-depending $C^*$-subalgebras of $\AB$.

\begin{definition}\label{valeleu}
We set
\begin{enumerate}
\item[(a)] $\BBA$ for the $C^*$-subalgebra of $\,\AB$ generated by
  $\S(\T)\equiv S^{-\infty}(\T) := \bigcap_{m \in \R} S^m_{\rho,\delta}(\T)$.
\item[(b)] $\MBA$ for the $C^*$-subalgebra of $\,\AB$ generated by
$S^0_{0,0}(\T)$.
\end{enumerate}
\end{definition}

It is easily observed that $\S(\T)$ is really independent of
$\rho$ and $\delta$. Part of our interest in the algebra $\BBA$ is due to the following
proposition and its corollary. We first recall that
\begin{equation*}
S^{-0}_{\rho,\delta}(\T):=\underset{m<0}{\bigcup}S^m_{\rho,\delta}(\T).
\end{equation*}

\begin{proposition}\label{infine}
For every $0\le\delta\le\rho\le 1$ with $\delta\ne 1$, the space
$S^{-0}_{\rho,\delta}(\T)$ is contained in $\BBA$.
\end{proposition}

\begin{proof}
We adapt the proof of Proposition~1.1.11 in \cite{Hoer} to show that any
$f \in  S^{-0}_{\rho,\delta}(\T)$ is the limit of a sequence
$ \bigl \{ f_{\epsilon} \bigr \}_{0 \leq \epsilon \leq 1} \in \S(\T)$ in the
topology of $S^0_{\rho,\delta}(\T)$, see also \cite[Sec.~1]{GS} for
more details. This and Proposition \ref{vaiancuru}
will imply the result.

Let $f \in S^m_{\rho,\delta}(\T)$ for some $m < 0$, $0 \leq
\delta \leq \rho \leq 1$,  $\delta \neq 1$, and let $\chi \in
\S(\X^*)$ with $\chi(0) = 1$. We set
$f_{\epsilon}(x,\xi):= \chi(\epsilon \xi)
\, f(x,\xi)$ for $0 \leq \epsilon \leq 1$.
By using Proposition~\ref{ionel}~(b), one has
$f_{\epsilon} \in \S(\T)$ for all $\epsilon >
0$, and $\bigl  \{ f_{\epsilon} \bigr \}_{0 \leq
\epsilon \leq 1}$ is a bounded subset of $S^m_{\rho,\delta}(\T)$.
Finally, one easily obtains that $f_{\epsilon}$ converges to $f$ as
$\epsilon \to 0$ in the topology of $S^{0}_{\rho,\delta}(\T)$.
\end{proof}

\begin{remark}
The same proof shows the density of $\S(\T)$ in $S^m_{\rho,\delta}(\T)$  with respect to the topology of
$S^{m'}_{\rho,\delta}(\T)$ for arbitrary $m'>m$.
\end{remark}

\begin{corollary}\label{labiel}
The $C^*$-algebra $\MBA$ is contained in the multiplier algebra $\M(\BBA)$ of $\BBA$.
\end{corollary}

\begin{proof}
This follows from the fact that $\S(\T)$ is a two-sided ideal in
$S^0_{0,0}(\T)$ with respect to $\sh$, from the definition of
$\BBA$ and $\MBA$ and from a density argument.
\end{proof}

Let us observe that ${\mathfrak B}^B_\C =C_0(\X^*)$ and ${\mathfrak M}^B_\C =BC_u  (\X^*)$,
while $\M({\mathfrak B}^B_\C) = BC(\X^*)$; so, in the Corollary, the inclusion could be strict.

\subsection{Magnetic twisted crossed products}\label{calll}

In the previous section we introduced some $C^*$-algebras through a
representation that was constructed with a vector potential $A$. However, all
these algebras did not depend on the choice of a particular $A$. Starting from
a magnetic twisted $C^*$-dynamical system, we shall now recall the
constructions of magnetic twisted $C^*$-algebras \cite{MPR1} and
relate them to the previous algebras.  These are
particular instances of twisted $C^*$-algebras extensively studied in
\cite{PR1} and \cite{PR2} (see also references therein).

We recall that Gelfand theory describes completely the structure of abelian
$C^*$-algebras. The Gelfand spectrum $\SA$ of $\A$ is the family of all
characters of $\A$ (a \textit{character} is just a morphism $\var:\A\rightarrow
\C$). With the topology of simple convergence $\SA$ is
a locally compact space, which is compact exactly when $\A$ is unital.

Since $\A\subset BC(\X)$, there exists a continuous surjection $\iota_\A:\beta(\X)\rightarrow
\SA$, where $\beta(\X)$ is the Stone-\v{C}ech compactification of the locally compact space $\X$.
By restriction, we get a continuous mapping with dense image (also denoted by $\iota_\A:\X\rightarrow
\SA$). This one is injective exactly when $C_0(\X)\subset\A$, case in which
$\SA$ is a compactification of $\X$.
The isomorphism between $\A$ and $C(\SA)$ can be precisely expressed as follows:
$\varphi:\X \to \C$ belongs to $\A$ if and only if there is a
(necessarily unique) $\tilde \varphi \in C(\SA)$ such that $\varphi=\tilde\varphi\circ\iota_\A$.
We shall extend the notation to functions depending on extra variables. For example, if $f:\Xi=\X\times\X^*\rightarrow\C$
is some convenient function, we define $\widetilde f:\SA\times\X^*\rightarrow\C$ by the property
$f(x,\xi)=\widetilde f(\iota_\A(x),\xi)$ for all $(x,\xi)\in\Xi$.

Let us finally mention that the map $\theta: \X \times \X \to \X,\ \theta(x,y):=x+y$
extends to a continuous map $\theta: \SA \times \X \to \SA$, because $\A$ was assumed
to be stable under translations. We also use the notations $\theta(\kappa,y)=\theta_y(\kappa)=\theta^\kappa(y)$
for $(\kappa,y)\in\SA\times\X$ and get a topological dynamical system $(\SA,\theta,\X)$ with compact space $\SA$.
Obviously one has $\iota_\A\circ\theta_y=\theta_y\circ\iota_\A$ for any $y\in\X$.

Now assume that the components $B_{jk}$ of the magnetic field belong to
$\A$. We define for each $x,y,z \in \X$ the expression
\begin{equation*}
\oB(x;y,z) := e^{- i \Gamma^B(\<x,x+y,x+y+z\>)}.
\end{equation*}
For fixed $x$ and $y$, the function $\oB(\cdot;x,y)\equiv \oB(x,y)$ belongs to the unitary group $\U(\A)$
of $\A$. Moreover, the mapping $\X \times \X \ni (x,y) \mapsto \oB(x,y) \in \U(\A)$
is a strictly continuous and normalized 2-cocycle on $\X$, i.e. for all $x,y,z \in \X$ the following
relations hold:
\begin{equation*}
\oB(x+y,z) \, \oB(x,y) = \theta_x[\oB(y,z)] \, \oB(x,y+z),
\end{equation*}
\begin{equation*}
\oB(x,0)=\oB(0,x)=1.
\end{equation*}
The quadruplet $(\A,\theta,\oB,\X)$ is a particular case of {\it a
twisted $C^*$-dynamical system} $(\A,\theta,\omega,\X)$. In the general case
$\X$ is a locally compact group, $\A$ is a $C^*$-algebra, $\theta$ is a continuous morphism from $\X$ to the group of
automorphisms of $\A$, and $\omega$ is a strictly continuous 2-cocycle with
values in the unitary group of the multiplier algebra of $\A$. We refer to
\cite[Def.~2.1]{MPR1} for more explanations.

Let $L^1(\X; \A)$ be the set of Bochner integrable functions on $\X$ with
values in $\A$, with the $L^1$-norm $\|F \|_1 :=\int_{\X} \de
x\;\!\|F(x)\|_\A$. For any $F,G \in L^1(\X;\A)$ and $x \in \X$, we define the
product
\begin{equation*}
(F \dia G)(x):=\int_{\X}\de y\;\theta_{\frac{y-x}{2}}\!\left[F(y)
\right]\;\!\theta_{\frac{y}{2}}\!\left[G(x-y)\right]\;\!\theta_{-\frac{x}{2}}\!\left[\oB(y,x-y)\right]
\end{equation*}
and the involution
\begin{equation*}
F^{\dia}(x):=\overline{F(-x)}.
\end{equation*}

\begin{definition}\label{primel}
The enveloping $C^*$-algebra of $L^1(\X;\A)$ is called {\rm the twisted crossed
product} and is denoted by $\A\!\rtimes^{\oB}_\theta\!\!\X$, or simply by $\CBA$.
\end{definition}

The $C^*$-algebras $\CBA$ and $\BBA$ are simply related by a partial Fourier transform
\begin{equation*}
[\FF(F)](\xi,x):=\int_{\X}\de y\,e^{iy\cdot\xi} F(y,x).
\end{equation*}

\begin{theorem}\label{ciudat}
The partial Fourier transform $\FF:\S'(\X\times \X)\rightarrow
\S'(\X^*\times \X)$ restricts to a $C^*$-isomorphism $\FF:\CBA\rightarrow\BBA$.
\end{theorem}

\begin{proof}
The partial Fourier transform $\FF$ is an isomorphism from $\S(\X;\A^\infty)$
to $\S(\T)$ which intertwines the products and the involutions:
\begin{equation*}
\FF(F)\sh \FF(G) = \FF[F\dia G],\ \ \ \ \ \big(\FF
(F)\big)^{\sh}=\FF\big(F^{\dia}\big).
\end{equation*}
The statement follows then from the density of $\S(\T)$ in $\BBA$, and from the
density of $\S(\X;\A^\infty)$ in $L^1(\X;\A)$, and hence also in $\CBA$.
\end{proof}

\begin{remark}
In Definition \ref{valeleu}, the algebra $\BBA$ was introduced as a closure
of a set of smooth elements, but it can easily be guessed that non-smooth elements
also belong to $\BBA$. Indeed, by \cite[Lemma A.4]{MPR2} one has that for any $m<0$ the set
$\FF^{-1}\big[S^{m}_1 (\X^*;\A )\big]$ is contained in $L^1(\X;\A)$,
which implies that $S^{m}_1 (\X^*;\A ) \subset \BBA$.
Here we have used the notation $S^{m}_1 (\X^*;\A )$ for the set of all functions
$f:\Xi \to \C$ that satisfy: (i) $f(\cdot,\xi) \in \A$ for all $\xi \in \X^*$,
(ii) $f(x,\cdot)\in C^\infty(\X^*)$ $\forall x\in \X$, and
(iii) for each $\alpha \in \N^n$ one has $\sigma_m^{\alpha0}(f)<\infty$.
Even more simply, one can also observe that the partial Fourier transform
of the elements in $L^1(\X;\A)$ belong to $\BBA$, and that these elements
do not necessarily possess any smoothness except continuity.
\end{remark}

\begin{remark}
In the same vein, let us also mention that a trivial extension of the same lemma
\cite[Lem.~A.4]{MPR2} to arbitrary $\delta$ imply that
$\FF^{-1}\big[S^{-0}_{1,\delta} (\X^*;\A^\infty)\big]$ is also contained in $L^1(\X;\A)$.
But by a remark in \cite[p.~17]{ABG} such an inclusion is no
longer true for $\rho\ne 1$. It follows that for $0\le\delta\le\rho<1$ many
elements of $\FF^{-1}\big[S^{-0}_{\rho,\delta}(\T)\big]$ only belong to
$\CBA\setminus L^1(\X;\A)$.
\end{remark}

We finally state a result about how the algebra $\BBA$
can be generated from a simpler set of its elements.
It is an adaptation of \cite[Prop.~2.6]{MPR1}.

\begin{proposition}\label{refraz}
The norm closure in $\AB$ of the subspaces generated either by $\{a\sh b\mid
a\in\A,\,b\in \S(\X^*)\}$ or by $\{b\sh a\mid b\in\A,\,b\in \S(\X^*)\}$ are
equal and coincide with the $C^*$-algebra $\BBA$.
\end{proposition}

We recall now the definition of a covariant representation of a magnetic
$C^*$-dynamical system. We denote by $\U(\H)$ the group of unitary operators in the Hilbert space $\H$.

\begin{definition}\label{RCT}
Given a magnetic $C^*$-dynamical system $(\A,\theta,\oB,\X)$, we call {\rm
covariant representation} $(\H,r,T)$ a Hilbert space $\H$ together with two
maps $r:\A\rightarrow \B(\H)$ and $T:\X\rightarrow \U(\H)$ satisfying
\begin{enumerate}[(a)]
\item $r$ is a non-degenerate representation,
\item $T$ is strongly continuous
and $\ T(x)\;\!T(y)=r[\oB(x,y)]\;\!T(x+y), \quad \forall x,y\in \X$,
\item $T(x)\;\!r(\varphi)\;\!T(x)^*=r[\theta_x(\varphi)], \quad \forall x\in \X,\;\varphi\in\A$.
\end{enumerate}
\end{definition}

\begin{lemma} If $(\H,r,T)$ is a covariant representation of
$(\A,\theta,\oB,\X)$, then $\Rep^T_r$ defined on $L^1(\X;\A)$ by
\begin{equation*}
\Rep_r^T(F):=\int_\X \de x\,r\left[\theta_{\frac{x}{2}}
\big(F(x)\big)\right]T(x)
\end{equation*}
extends to a representation of $\CBA=\A\rtimes^{\oB}_\theta\!\!\X$.
\end{lemma}

One can recover the covariant representation from $\Rep^T_r$.
Actually, there is a one-to-one correspondence between covariant
representations of a twisted $C^*$-dynamical system and non-degenerate
representations of the twisted crossed product, which preserves
unitary equivalence, irreducibility and direct sums.

By composing with the partial Fourier transformation, one gets the most general
representations of the pseudodifferential $C^*$-algebra $\BBA$, denoted by
\begin{equation*}
\Op^T_r:\BBA\rightarrow\B(\H),\ \ \ \ \ \Op^T_r(f):=\Rep^T_r[\FF^{-1}(f)].
\end{equation*}
Given any continuous vector potential $A$ we construct a
representation of $\CBA$ in $\H=L^2(\X)$. For any $u \in \H$ and $x,y \in \X$,
we define {\it the magnetic translations}
\begin{equation*}
[T^A(y)u](x):=\lambda^A(x;y)\;\!u(x+y) = e^{-i\Gamma^A([x,x+y])}\;\!u(x+y).
\end{equation*}
Let us also set $r(\varphi):=\varphi(Q)$ for any $\varphi \in
\A$, where $\varphi(Q)$ denotes an operator of multiplication in $\H$. Then the
triple $(\H,r,T^A)$ is a covariant representation of the magnetic
$C^*$-dynamical system, see \cite[sec.~4]{MPR1} for details. The
corresponding representation $\Rep_r^{T^A}$ of the algebra $\CBA$, denoted by
$\Rep^A$, is explicitly given for any $F\in L^1(\X;\A)$ and any $u \in \H$
by
\begin{equation}\label{representation}
\big[\Rep^A(F)u\big](x) = \int_\X \de
y\;\!\lambda^A(x;y-x)\;\!F\big(\hbox{$\frac{1}{2}$}(x+y);y-x\big)\;\!u(y).
\end{equation}
This representation is called \emph{the Schr\"odinger representation of $\CBA$
associated with the vector potential} $A$. It is proved in \cite[Prop.~2.17]{MPR1} that this
representation is faithful.
We recall that the choice of another vector potential generating
the same magnetic field would lead to a unitarily equivalent representation of
$\CBA$ in $\B(\H)$. By comparing \eqref{op} and \eqref{representation}, one
sees that $\Op^A\equiv\Op^{T^A}_r$ and $\Rep^A$ are connected by the partial Fourier transform:
$\Op^A(f) = \Rep^A[\FF^{-1}(f)]$ for suitable $f$.

\subsection{Inversion}\label{versi}

The following approach is mainly inspired by a similar
construction in \cite[Sec.~7.1]{IMP2}. It also relies on some basic results on
$\Psi^*$-algebras that we borrow from \cite[Sec.~2]{Lau}, see also \cite{LMN}
and references therein.

Let $\CCC$ be a $C^*$-algebra with unit $1$, and let $\SS$ be a
$^*$-subalgebra of $\CCC$ with $1 \in \SS$. $\SS$ is called
\emph{spectrally invariant}  if $\SS \cap
\CCC^{-1}=\SS^{-1}$, where $\SS^{-1}$, resp.~$\CCC^{-1}$,
denotes the set of invertible elements in $\SS$, resp.~$\CCC$.
Furthermore, $\SS$ is called a \emph{$\Psi^*$-algebra} if it is
spectrally invariant and endowed with a Fr\'echet topology such that the
embedding $\SS\hookrightarrow\CCC$ is continuous. It is shown in
\cite[Cor.~2.5]{Lau} that a closed $^*$-subalgebra of a $\Psi^*$-algebra (also
containing $1$), endowed with the restricted topology, is also a
$\Psi^*$-algebra.

It has been proved in \cite{IMP2} that for $\rho \in [0,1]$,
$S^0_{\rho,0}(\Xi)$  is a $\Psi^*$-algebra in $\AB$.
Then, since $S^0_{\rho,0}(\T)$ is a closed $^*$-subalgebra of
$S^0_{\rho,0}(\Xi)$ by our Lemma \ref{ionel} (a) and Theorem \ref{babana}, it
follows that $S^0_{\rho,0}(\T)$ is a $\Psi^*$-algebra in $\AB$. In particular,
if $f \in S^0_{\rho,0}(\T)$ has an inverse in $\AB$ with respect to $\sh$, denoted by $f^{(-1)_B}$,
then this inverse belongs to $S^0_{\rho,0}(\T)$.
As by-products of the theory of
$\Psi^*$-algebras, one can state
\begin{proposition}
$S^0_{\rho,0}(\T)$ is a $\Psi^*$-algebra, it is stable under the holomorphic
functional calculus, $[S^0_{\rho,0}(\T)]^{(-1)_B}$ is open and the map
$[S^0_{\rho,0}(\T)]^{(-1)_B}\ni f\mapsto f^{(-1)_B}\in S^0_{\rho,0}(\T)$ is
continuous. 
\end{proposition}

In order to state the next lemma some notations are needed. For
$m>0$, $\lambda>0$ and $\xi \in \X^*$, set
\begin{equation*}
\p_{m,\lambda}(\xi):=\<\xi\>^m+\lambda.
\end{equation*}
The map $\p_{m,\lambda}$ is clearly an element of $S^m_{1,0}(\T)$ and
its pointwise inverse an element of $S^{-m}_{1,0}(\T)$. It has been proved in \cite[Thm.~1.8]{MPR2}
that for $\lambda$ large enough, $\p_{m,\lambda}$ is invertible with respect to
the composition law $\sh$ and that its inverse $\p_{m,\lambda}^{(-1)_B}$
belongs to $\BBA$. So for any $m>0$ we can fix $\lambda=\lambda(m)$ such that
$\p_{m,\lambda(m)}$ is invertible. Then, for arbitrary $m \in \R$, we set
\begin{equation*}
\r_m:=\left\{\begin{array}{ll} \p_{m,\lambda(m)} & \hbox{ for } m>0 \\
\p_{|m|,\lambda(|m|)}^{(-1)_B} & \hbox{ for } m<0 \end{array}\right.
\end{equation*}
and $\r_0:=1$. The relation $\r_m^{(-1)_B}=\r_{-m}$ clearly holds for all $m
\in \R$.  Let us still show an important property of $\r_m$.

\begin{lemma}
For any $m \in \R$, one has $\r_m\in S^m_{1,0}(\T)$.
\end{lemma}

\begin{proof}
For $m\geq 0$, the statement is trivial from the definition of $\r_m$.
But for $m<0$ the function $\r_m$ will also depend on the variable $x$, so
one has to take the proof of \cite[Thm.~1.8]{MPR2} into account.
Indeed, it has been shown in that reference that for $\lambda$ large enough,
$\p:=\p_{|m|,\lambda(|m|)}$ is invertible with respect to the composition law
$\sh$, and that its inverse is given by the formula
\begin{equation}\label{petitjeu}
\p^{(-1)_B}= \p^{-1} \sh (\p \sh \p^{-1})^{(-1)_B},
\end{equation}
where $\p^{-1}$ is the the inverse of $\p$ with respect to pointwise
multiplication, and $\lambda$ has been chosen such that $(\p \sh
\p^{-1})^{(-1)_B}$ is well defined and belongs to $\AB$. Furthermore, since
$\p^{-1}$ belongs to $S^{-m}_{1,0}(\T)$, the product $\p \sh \p^{-1}$ belongs
to $S^0_{1,0}(\T)$. It then follows that the inverse of $\p \sh \p^{-1}$ also
belongs to $S^0_{1,0}(\T)$ by the $\Psi^*$-property of $S^0_{1,0}(\T)$. One
concludes by observing that the r.h.s.~of \eqref{petitjeu} belongs to
$S^{-m}_{1,0}(\T)$, and corresponds to $\r_m$ for $m<0$.
\end{proof}

\begin{proposition}\label{lolo}
Let $m>0$, $\rho \in [0,1]$ and $f \in S^m_{\rho,0}(\T)$. If $f$ is
invertible in $\MB(\Xi)$ and $\r_m \sh f^{(-1)_B} \in \AB$, then
$f^{(-1)_B}$ belongs to $S^{-m}_{\rho,0}(\T)$.
\end{proposition}

\begin{proof}
Let us first observe that
\begin{equation*}
f\sh \r_{-m} \in  S^m_{\rho,0}(\T) \;\sh\; S^{-m}_{1,0}(\T)\subset
S^0_{\rho,0}(\T).
\end{equation*}
Furthermore, this element is invertible in $\AB$ since its inverse $\big(f\sh
\r_{-m}\big)^{(-1)_B}$ is equal to $\r_m \sh f^{(-1)_B}$, which belongs to
$\AB$. Then, by the $\Psi^*$-property of $S^0_{\rho,0}(\T)$, it
follows that  $\big(f\sh \r_{-m}\big)^{(-1)_B}$ belongs to $S^0_{\rho,0}(\T)$,
and so does $\r_m \sh f^{(-1)_B}$. Consequently, one has
\begin{equation*}
f^{(-1)_B} = \r_{-m} \sh [\r_m \sh f^{(-1)_B}] \in S^{-m}_{\rho,0}(\T) \; \sh
\; S^0_{\rho,0}(\T) \subset S^{-m}_{\rho,0}(\T).
\end{equation*}
\end{proof}

In order to verify the hypotheses of the above proposition, a condition of
ellipticity is usually needed.

\begin{definition}
A symbol $f \in S^m_{\rho,\delta}(\T)$ is called \emph{elliptic} if there exist $R,C>0$ such that
\begin{equation*}
|f(x,\xi)|\geq C \<\xi\>^m
\end{equation*}
for all $x \in \X$ and $|\xi|>R$.
\end{definition}

We are now in a position to state and prove our main theorem on inversion (see also the Appendix):

\begin{theorem}\label{version}
Let $m>0$, $\rho \in [0,1]$ and $f$ be a real-valued elliptic element of
$S^m_{\rho,0}(\T)$. Then for any $\z \in\C\setminus\R$ the function
$f-\z$ is invertible in $\MB(\Xi)$ and its
inverse $(f-\z)^{(-1)_B}$ belongs to $S^{-m}_{\rho,0}(\T)$.
\end {theorem}

\begin{proof}
It has been proved in \cite[Thm.~4.1]{IMP1} that $\Op^A(f)$
defines a self-adjoint operator in $\H:=L^2(\X)$ for any vector potential
$A $ whose components belong to $C^\infty_{\rm{pol}}(\X)$. In
particular, $\z$ does not belong to the spectrum of $\Op^A(f)$, which is
independent of $A$ by gauge covariance, and
$\Op^A(f)-\z = \Op^A(f-\z)$ is invertible. Its inverse belongs to
$\B(\H)$, which means that $(f-\z)^{(-1)_B}$ exists in $\MB(\Xi)$ and belongs to
$\AB$. Furthermore, Theorem 4.1 of \cite{IMP1} also implies that
$\Op^A\big[(f-\z) \sh \r_m^{(-1)_B}\big]$ is a bijection on $\H$, and
thus $\r_m\sh (f-\z)^{(-1)_B}=\big[(f-\z)\sh \r_m^{(-1)_B}\big]^{(-1)_B} \in \AB$.
One finally concludes by taking Proposition \ref{lolo} into account.
\end{proof}

\subsection{Affiliation}\label{mmrr}

We start by recalling the meaning of affiliation, borrowed from
\cite{ABG}. We shall then prove that some of the classes of symbols introduced
in Section \ref{secintro} define observables affiliated to $\BBA$.

\begin{definition}\label{secundel}
\emph{An observable affiliated to a $C^*$-algebra} $\CCC$ is a morphism $\Phi:
C_0(\R) \to \CCC$.
\end{definition}

For example, if $\H$ is a Hilbert space and $\CCC$ is a $C^*$-subalgebra
of $\B(\H)$, then a self-adjoint operator $H$ in $\H$ defines an observable
$\Phi_{\!\hbox{\it \tiny H}}$ affiliated to $\CCC$ if and only if
$\Phi_{\!\hbox{\it \tiny H}}(\eta) := \eta(H)$ belongs to $\CCC$ for all $\eta
\in C_0(\R)$. A sufficient condition is that $(H-\z)^{-1} \in \CCC$ for some $\z
\in \C$ with $\im \z \neq 0$. Thus an observable affiliated to a $C^*$-algebra
is the abstract version of the functional calculus of a self-adjoint operator.

The next result is a rather simple corollary of our previous results. We call
it Theorem to stress its importance in our subsequent spectral results.

\begin{theorem}\label{afileisn}
For $m>0$ and $\rho \in [0,1]$, any real-valued elliptic element $f\in
S^m_{\rho,0}(\T)$ defines an observable affiliated to the $C^*$-algebra $\BBA$.
\end{theorem}

\begin{proof}
For $\z\in \C\setminus\R$, let us set $r_\z:=(f-\z)^{-1}$.
We also define $\Phi(r_\z):=(f-\z)^{(-1)_B}$.
We first prove that the family $\{\Phi(r_\z)\mid \z \in\C\setminus \R \}$
satisfies the resolvent equation.
Indeed, for any two complex numbers $\z,\z' \in \C\setminus \R$, one has
\begin{equation*}
(f-\z)\sh \Phi(r_\z)=1 \qquad \hbox{ and }\qquad (f-\z')\sh \Phi(r_{\z'})=1.
\end{equation*}
By subtraction, one obtains
$(f-\z)\sh[\Phi(r_\z)-\Phi(r_{\z'})] + (\z'-\z)\Phi(r_{\z'})=0$.
By multiplication to the left with $\Phi(r_\z)$ and using the associativity,
one then gets the resolvent equation
\begin{equation*}
\Phi(r_\z)-\Phi(r_{\z'}) = (\z-\z')\;\!\Phi(r_\z)\sh\Phi(r_{\z'}) \ .
\end{equation*}

We has thus obtained a map $\C\setminus \R \ni \z \mapsto \Phi(r_\z) \in
S^{-m}_{\rho,0}(\T)\subset \BBA$, where Theorem \ref{version} and Proposition
\ref{infine} have been taken into account. Furthermore, the relation
$\Phi(r_\z)^{\sh}=\Phi(r_{\overline{\z}})$ clearly holds.
A general argument presented in \cite[p.~364]{ABG} allows now to extend in a unique
way the map $\Phi$ to a $C^*$-algebra morphism $C_0(\R)\to\BBA$.
\end{proof}

\section{Spectral analysis}\label{secspectral}

\subsection{Preliminaries}\label{esenici}

Recall that $\A$ is a $C^*$-subalgebra of $BC_u(\X)$ which is invariant under translations.
Such an algebra is called \emph{admissible}.
It is also unital, but most of the constructions do not require this.
For any $\varphi \in \A$ we systematically denote
by $\tilde \varphi$ the unique element
of $C(\SA)$ satisfying $\varphi=\tilde\varphi\circ\iota_\A$. In fact,
$\tilde \varphi$ corresponds to the image of $\varphi$ through the Gelfand isomorphism
$\G_\A:\A\rightarrow C(\SA)$.

A basic fact is that $\A$ comes together with a family of short exact sequences
\begin{equation}\label{short}
0\longrightarrow\A^\Q\longrightarrow\A\overset{\pi_\Q}
{\longrightarrow}\A_\Q\longrightarrow 0
\end{equation}
indexed by $\mathbf Q_\A$, the set of all quasi-orbits of the topological dynamical
space $(\SA,\theta,\X)$. We recall that a quasi-orbit is the closure of an orbit, and let
us now explain the meaning of \eqref{short}.

For $\Q\in\mathbf Q_\A$ we say that the element $\var\in \SA$ \emph{generates} $\Q$ 
if the orbit of $\var$ is dense in $\Q$.
In general not all the elements of $\Q$ generates it.
There is a canonical epimorphism $p_\Q:C(\SA)\rightarrow C(\Q)$, coming from the inclusion of
the closed set $\Q$ in $\SA$. On the other hand, if $\var$ generates $\Q$, we set
$\A_\var:=\{\varphi_\var:=\tilde\varphi\circ\theta^\var \mid\widetilde\varphi\in C(\Q)\}$.
It is clear that $\A_\var$ is an admissible $C^*$-algebra isomorphic to $C(\Q)$.
Indeed, by taking into account the surjectivity of the
morphism $p_\Q$ and the continuity of translations in $\A \subset
BC_u(\X)$, one easily sees that $\varphi_{\var}: \X \to \C$ belongs to $BC_u(\X)$.
Furthermore, the induced action of $\X$ on $\varphi_{\var}$ coincides with the
natural action of $\X$ on $BC_u(\X)$.
Thus, we get an epimorphism $\pi_\var:\A\rightarrow\A_\var$ by
$\pi_\var:=\theta^\var\circ p_\Q\circ\G_\A$.
We note that in general $\A_\var$ has no reason to be contained in $\A$.

It is clear that the kernel of this epimorphism is
$\A^\Q=\{\varphi\in\A\mid \tilde\varphi|_\Q=0\}$.
Furthermore, if $\var$ and $\var'$ generate the same quasi-orbit $\Q$, the algebras
$\A_\var$ and $\A_{\var'}$ are isomorphic. So by a slight abuse of notation, we
call them generically $\A_\Q$, and corresponding morphism by $\pi_\Q$ .
This finishes to explain \eqref{short}.

We now recall some more definitions in relation with spectral analysis in a
$C^*$-algebraic framework, {\it cf.}~\cite{ABG}.
Let $\Phi$ be an observable affiliated to a $C^*$-algebra $\CCC$ and let
$\KK$ be an ideal of $\CCC$. Then, {\it the $\KK$-essential spectrum of
$\Phi$} is
\begin{equation*}
\sigma_\KK(\Phi):= \big\{\lambda \in \R\;|\; \hbox{ if }\eta \in C_0(\R) \hbox{
and } \eta(\lambda) \neq 0, \hbox{ then }\Phi(\eta) \not \in \KK\big\}.
\end{equation*}
If $\pi$ denotes the canonical morphism $\CCC \to \CCC/\KK$, 
then $\pi[\Phi]:C_0(\R)\to \CCC/\KK$ given by 
$\big(\pi[\Phi]\big)(\eta):=\pi[\Phi(\eta)]$ is an
observable affiliated to the quotient algebra, and one has
$\sigma_\KK(\Phi) = \sigma_{\{0\}}(\pi[\Phi])\equiv\sigma(\pi[\Phi])$.
Assume now that $\CCC$ is a $C^*$-subalgebra of $\B(\H)$ for
some Hilbert space $\H$ and that $\CCC$ contains the ideal $\K(\H)$ of
compact operators on $\H$. Furthermore, let $H$ be a self-adjoint operator
in $\H$ affiliated to $\CCC$. Then $\sigma_{\K(\H)}(\Phi_{\!\hbox{\it \tiny H}})$
is equal to the essential spectrum $\sigma_{\hbox{\rm \tiny ess}}(H)$ of $H$.
Here we shall be mainly interested in the usual spectrum and in the essential
spectrum. The need for the $\KK$-essential spectrum with $\KK$ different from
$\{0\}$ or $\K(\H)$ is relevant in the context of Remark \ref{nonprop}.

\subsection{The essential spectrum of anisotropic magnetic operators}\label{esentz}

We consider again the magnetic twisted $C^*$-dynamical
system $(\A, \theta, \oB, \X)$ and explain how to calculate the essential
spectrum of any observable affiliated to the twisted crossed product algebra
$\CBA$. Then, by using the results of the previous sections, we particularize to the case of
magnetic pseudodifferential operators and prove our main result concerning
their essential spectrum. For simplicity, we omit in this section the subscript $B$ 
to the $2$-cocycle $\oB$.

We follow now the strategy of \cite{M1,MPR2} (see also references therein).
We are going to assume that $\A$ contains $C_0(\X)$, so $\SA$
is a compact space and $\X$ can be identified with a dense open subset of $\SA$.
Since the group law $\theta: \X \times \X \to \X$ extends to a continuous map $\theta:
\X \times \SA \to \SA$, the complement $\FA$ of $\X$ in $\SA$ is closed and
invariant; it is the space of a compact topological dynamical system.
For any quasi-orbit $\Q$, the algebra $\A^\Q$ is clearly  an invariant ideal of $\A$.
The abelian twisted dynamical system $(\A^\Q, \theta,\omega, \X)$ obtained by replacing $\A$ with
$\A^\Q$ and performing suitable restrictions is well defined. Furthermore, the
twisted crossed product $\A^\Q\!\!\rtimes^{\omega}_\theta\!\X$ may be identified
with an ideal of $\A\!\rtimes^{\omega}_\theta\!\X$ \cite[Prop.~2.2]{PR2}.

In order to have an explicit description of the quotient, let us first note
that $\A / \A^\Q$ is canonically isomorphic to the unital $C^*$-algebra $C(\Q)$
of all continuous functions on $\Q$. The natural action of $\X$ on $\tilde \varphi \in C(\Q)$
is given by $(\theta_x \tilde \varphi)(\var) = \tilde \varphi\left(\theta_x[\var]\right)$
for each $x \in \X$ and $\var \in \Q$. Now, the restriction of $\omega$ to $\Q$
gives rise to a 2-cocycle $\omega_\Q: \X \times \X \to \U\big(C(\Q)\big)$
precisely defined by $\omega_\Q(x,y):= p_\Q\big[\G_\A\big(\omega(x,y)\big)\big]$ for each $x,y\in\X$.
Thus $\big(C(\Q), \theta,
\omega_\Q,\X\big)$ is a well-defined abelian twisted $C^*$-dynamical
system. Moreover, the quotient $\A\!\rtimes^{\omega}_\theta\!\X /
\A^\Q\!\!\rtimes^{\omega}_\theta\!\X$ may be identified with the corresponding
twisted crossed product $C(\Q)\!\rtimes^{\omega_{\!\Q}}_\theta\!\X$. This follows
from \cite[Prop.~2.2]{PR2} if $\A$ is separable. For the non-separable case,
just perform obvious modifications  in the proof of \cite[Th.~2.10]{GI3} to
accommodate the $2$-cocycle.
Finally, by taking the isomorphisms $\pi_\Q$ introduced in Section \ref{esenici} into account,
the algebra $C(\Q)\!\rtimes^{\omega_{\!\Q}}_\theta\!\X$ is isomorphic to
$\A_\Q\!\rtimes^{\omega_{\!\Q}}_\theta\!\X$, and the canonical morphism
$\Pi_\Q:\A\!\rtimes^{\omega}_\theta\!\X  \to
\A_\Q\!\rtimes^{\omega_\Q}_\theta\!\X$ is defined on any $F \in L^1(\X;\A)$
by $\big(\Pi_\Q[F]\big)(x) = \pi_\Q\big(F(x)\big)$ for all $x \in \X$.

Let us now consider $\mathbf Q\subset \mathbf Q_\A$ such that
the elements $\Q$ of $\mathbf Q$ define a covering of $\FA$.
At the algebraic level, the covering requirement reads $\cap_{\Q \in \mathbf Q}
\A^\Q=C_0(\X)$. This implies immediately the equality
\begin{equation*}
\bigcap_{\Q \in \mathbf Q} \; \A^\Q\!\!\rtimes^{\omega}_\theta \!\X =
C_0(\X)\!\rtimes^{\omega}_\theta\!\X\;.
\end{equation*}
By putting all these together one obtains, {\it cf.}~also \cite[Prop.~1.5]{M1}:

\begin{proposition}\label{Propess}
Let $\mathbf Q\subset \mathbf Q_\A$ define a covering of $\FA$ by quasi-orbits.
\begin{enumerate}
\item[\rm{(a)}] There exists an injective morphism:
$\A\!\rtimes^{\omega}_\theta\!\X \ / \ C_0(\X)\!\rtimes^{\omega}_\theta\!X
\hookrightarrow \prod_{\Q \in \mathbf Q}
\A_\Q\!\rtimes^{\omega_{\Q}}_\theta\!\X$,
\item[\rm{(b)}] If $\Phi$ is an observable affiliated to
$\A\!\rtimes^{\omega}_\theta\!\X$ and $\KK:=
C_0(\X)\!\rtimes^{\omega}_\theta\!\X$, then we have
\begin{equation}\label{spectre}
\sigma_{\KK}(\Phi) = \overline{\bigcup_{\Q \in \mathbf Q} \sigma(\Pi_\Q[\Phi])}\;.
\end{equation}
\end{enumerate}
\end{proposition}

We now introduce a represented version of this proposition in the Hilbert space
$\H:=L^2(\X)$. We recall that for any continuous vector potential $A$, a representation
$\Rep^A$ of $\A\!\rtimes^{\omega}_\theta\!\X$ has been introduced in
\eqref{representation}, and that this representation is irreducible and
faithful \cite[Prop.~2.17]{MPR1}. Furthermore, it is proved in the same
reference that $\Rep^A\big(C_0(\X)\!\rtimes^{\omega}_\theta\!\X\big)$ is equal
to $\K(\H)$. If $\Phi$ is an observable affiliated to
$\A\!\rtimes^\omega_\theta\!X$, then the l.h.s.~term of \eqref{spectre} is
equal to $\sigma_{\hbox{\rm \tiny ess}}\big(\Rep^A(\Phi)\big)$, and it does not
depend on a choice of $A$.

We are now in a position to prove a concrete result for the calculation of the
essential spectrum of any magnetic pseudodifferential operator. It consists
essentially in an application of Proposition \ref{Propess} together with a
partial Fourier transformation. It also relies on the affiliation result obtained in 
Theorem \ref{afileisn}.
The components of the magnetic field $B_\Q$ are defined by $\pi_\Q(B_{jk})$.

\begin{theorem}\label{thmess}
Let $m>0$, $\rho \in [0,1]$ and let $\mathbf Q\subset \mathbf Q_\A$ define a covering of $\FA$.
Then, for any real-valued elliptic element $f$ of $S^m_{\rho,0}(\T)$ one has
\begin{equation*}
\sigma_{\hbox{\rm \tiny ess}}\big[\Op^A(f)\big] = \overline{\bigcup_{\Q \in \mathbf Q}
\sigma[\Op^{A_\Q}(f_\Q)]},
\end{equation*}
where $A$ and $A_\Q$ are continuous vector potentials for $B$ and $B_\Q$,
and $f_\Q \in S^m_{\rho,0}\big(\X^*;\A_\Q\big)$ is the image of $f$ through $\pi_\Q$.
\end{theorem}

\begin{proof}
Let us first observe that the morphism
\begin{equation*}
\FF\big(L^1(X;\A)\big) \ni g \mapsto \FF\big(\Pi_{\Q}[\FF^{-1}(g)]\big) \in
\FF\big[L^1\big(X;C(\Q)\big)\big]
\end{equation*}
extends to a surjective morphism $\tilde\Pi_{\Q}:\BBA \to
{\mathfrak B}^{B_\Q}_{\A_\Q}$. The equality \eqref{spectre} can
then be rewritten in this framework and for any observable $f$ affiliated to $\BBA$:
\begin{equation*}
\sigma_{\KK}(f) = \overline{\bigcup_{\Q \in \mathbf Q}
\sigma\big(\tilde\Pi_{\Q}[f]\big)},
\end{equation*}
where $\KK$ is now the ideal of $\BBA$ given by the image of
$C_0(\X)\!\rtimes^{\omega}_\theta\!\X$ through the map $\FF$.
The result follows now from the central observation that
$\tilde\Pi_{\Q}[f]$ is equal to $f_{\Q}$ and by considering faithful
representations of $\BBA$ through $\Op^{A}$ and of ${\mathfrak
B}^{B_\Q}_{\A_\Q}$ through $\Op^{A_\Q}$.
\end{proof}

\begin{remark}\label{prak}
Combining our approach with techniques from \cite{ABG,GI1,GI3}, one could extend the result above to more singular symbols $f$.
We shall not do this; our main goal was to cover functions $f$ which have no specific dependence of the variable in $\Xi$
(as $f(x,\xi)=h(\xi)+V(x)$ for instance) in a pseudodifferential setting.
\end{remark}

\begin{remark}\label{prac}
To have a good understanding of (\ref{thmess}), one needs admissible algebras $\A$ for which the space $\mathbf Q_\A$
is explicit enough. Many examples are scattered through \cite{ABG,AMP,GI1,GI2,GI3,M1,MPR2,Ric}
and we will not reconsider this topic here.
\end{remark}

\begin{remark}\label{nonprop}
Non-propagation results easily follow from this algebraic framework. They have
been explicitly exhibited in the non-magnetic case in \cite{AMP} and in the
magnetic case in \cite{MPR2}. In these references, the authors were mainly
concerned with generalized Schr\"odinger operators and their results were
stated for these operators. But the proof relies only on the $C^*$-algebraic
framework, and the results extend mutatis mutandis to the classes of symbols
introduced in the present paper. For shortness, we do not present these
propagation estimates here, but statements and proofs can easily be mimicked from these references.
\end{remark}

\section*{Appendix: An independent proof for the affiliation}

In this Appendix, we give a second proof of Theorem \ref{afileisn} which is independent 
of the results contained in \cite{IMP2}, which is not yet published.

Let $m>0$, $\rho \in (0,1]$ and $f$ be a real-valued elliptic element of
$S^m_{\rho,0}(\T)$. For some $\z \in \C$, we are first going to show that
$(f-\z)^{(-1)_B}$ belongs to $\BBA$ by writing
down a series for the inverse $(f-\z)^{(-1)_B}$ of the form
\begin{equation*}
(f-\z)^{(-1)_B}=(f-\z)^{-1}\sh\sum_{k=0}^\infty \left[1-(f-\z)\sh (f-\z)^{-1}\right]^{k\sh},
\end{equation*}
with $(f-\z)^{-1}$ the pointwise inverse of $f-\z$. Notice that $(f-\z)^{-1}$ belongs to 
$S^{-m}_{\rho,0}(\T)\subset\BBA$ by ellipticity, and that $g^{k\sh}$ denotes the $k$'th power of $g$ with respect to $\sh$.
By the asymptotic development, one knows that the reminder $R_\z:=(f-\z)\sh
(f-\z)^{-1}-1$ belongs to $S^{-\rho}_{\rho,0}(\T)\subset \BBA$. However, an
additional argument is needed to show that the $\|\cdot\|_B$-norm of $R_\z$ is
subunitary for suitable $\z$, insuring the convergence of the Neumann series.

For that purpose, we recall from Section \ref{brasagain} that  $\| R_\z\|_B\,\equiv\,\| R_\z\|_{\AB}
:=\| \Op^A(R_\z)\|_{\B(\H)}$. Furthermore, from the magnetic Calderon-Vaillancourt
theorem this norm can be estimated from above with expressions of the form
\begin{equation}\label{zdependent}
\sup_{(x,\xi) \in \Xi} \expval{\xi}^{\rho(|\delta|-|d|)} \babs{\partial_x^d
\partial_{\xi}^{\delta} R_\z(x,\xi)}
\end{equation}
for a finite number of multi-indices $\delta,d \in \N^n$, see Theorem \ref{uru} for the precise statement.
Thus it remains to study the dependence on $\z$ of \eqref{zdependent}. Fortunately,
a similar expression has already been studied in the proof of the asymptotic development
and we shall rely on some of the expressions derived in the proof of Theorem \ref{bibigul}.

Since $f$ is an elliptic symbol of strictly positive order we can fix $\z \in \R_-$
with $\z\leq \inf f-1$. The pointwise inverse of $f-\z$ is thus well defined,
and is denoted by $(f-\z)^{-1}$. We recall from the proof of Theorem \ref{bibigul} that
\begin{equation}\label{defofR}
R_\z(X)=\underset{|a|+|b|+|\alpha|+|\beta|=1}{\underset{\alpha'\le\alpha,\,
\beta'\le\beta}{\underset{a,b,\alpha,
\beta,\alpha',\beta'}{\sum}}}\int^1_0 \de \tau\,{\rm
pol}_{\a,\b}^{\alpha',\beta'} (\tau)\  I^{\alpha',\beta'}_{\tau,\z,\a,\b}(X),
\end{equation}
where ${\rm pol}_{\a,\b}^{\alpha',\beta'} :[0,1]\to\C$ are polynomials and
\begin{eqnarray*}
I^{\alpha',\beta'}_{\tau,\z,\a,\b}(X):=&& \int_\X \de y \int_\X \de z
\int_{\X^*} \de \eta \int_{\X^*} \de \zeta \,e^{-2i\sigma(Y,Z)}\;
[\partial^{\alpha-\alpha'}_z\partial^{\beta-\beta'}_y\ob]
(x,y,z)\ \cdot \\
&&\cdot\ [\partial^{a+\beta'}_x\partial^{\alpha+b}_\xi (f-\z)](x-\tau
y,\xi-\tau\eta)\; [\partial^{b+\alpha'}_x
\partial^{a+\beta}_\xi (f-\z)^{-1}] (x-\tau z,\xi-\tau\zeta).
\end{eqnarray*}
Retaining only its essential features, we shall rewrite this last
expression as
\begin{equation*}
I_{\tau,\z}(X):=\int_\X \de y \int_\X \de z \int_{\X^*} \de \eta
\int_{\X^*} \de \zeta  \,e^{-2i\sigma(Y,Z)}\;\sb(x,y,z)\;F_\z(x-\tau
y,\xi-\tau\eta) \;G_\z(x-\tau z,\xi-\tau\zeta)\ .
\end{equation*}

In order to obtain estimates for \eqref{zdependent}, let us
calculate
$\partial^{d}_{x}\partial^{\delta}_{\xi}I_{\tau,\z}$. Actually, by
using \eqref{froufrou}, the oscillatory integral
definition of $\partial_x^d\partial_\xi^\delta I_{\tau,\z}$ is
\begin{equation*}
[\partial^{d}_{x}\partial^{\delta}_{\xi}I_{\tau,\z}](X)
=\,\underset{\delta^1+\delta^2=\delta}{\underset{d^0+d^1+d^2=d}{\sum}}
C^{\delta^1 \delta^2}_{d^0 d^1 d^2} \int_\X \de y \int_\X \de z
\int_{\X^*} \de \eta \int_{\X^*} \de \zeta\; e^{-2i\sigma(Y,Z)}
\;L^{\tau,\z, \delta^1, \delta^2}_{p, q, d^0, d^1, d^2}(X,Y,Z)\ ,
\end{equation*}
where, for suitable integers $p,q$, the expression $L^{\tau,\z, \delta^1,
\delta^2}_{p, q, d^0, d^1, d^2}(X,Y,Z)$ is given by
\begin{eqnarray*}
&& \<\eta\>^{-2p}\<\zeta\>^{-2p}\<y\>^{-2q}\<z\>^{-2q}\underset{|q^1|\leq
q,\,|q^2|\leq q}{\underset{|c^1|+|c^2|+|c^3|=2p}{\underset{|b^1|+|b^2|+ |b^3|=2p}{\sum}}}
C^{q^1 q^2 c^1 c^2 c^3}_{b^1 b^2 b^3} \varphi_{q c^1}(z)\;\! \psi_{q
b^1}(y)\;\!  [\partial^{d^0}_x\partial^{b^2}_y\partial^{c^2}_z\sb](x,y,z)\ \cdot \\
&\cdot &(-\tau)^{2|q^1|+2|q^2|+|b^3|+|c^3|}
[\partial^{d^1+b^3}_x\partial^{\delta^1+2q^1}_\xi F_\z]
(x-\tau y,\xi-\tau\eta) \;
[\partial^{d^2+c^3}_x\partial^{\delta^2+2q^2}_\xi G_\z](x-\tau z,\xi-\tau\zeta),
\end{eqnarray*}
where $\varphi_{q c^1}$ and $\psi_{q b^1}$ are bounded
functions produced by derivating the factors $\<z\>^{-2q}$ and
$\<y\>^{-2q}$, respectively.
We need now an explicit dependence on $\z$ of the last two factors.

Let us first recall that $F_\z=\partial^{a+\beta'}_x\partial^{\alpha+b}_\xi (f-\z)$,
and hence two distinct situations occur: If $a=\beta'=\alpha=b=d^1=b^3=\delta^1=q^1=0$,
then
\begin{equation*}
\big|[\partial^{d^1+b^3}_x\partial^{\delta^1+2q^1}_\xi F_\z]
(x-\tau y,\xi-\tau\eta) \big| \equiv |f(x-\tau y,\xi-\tau\eta)-\z|
\end{equation*}
and this is the annoying contribution that has to be dealt separately below. But if any of the above multi-indices is
non-null, then the dependence on $\z$ vanishes, and one has
\begin{equation*}
\big|[\partial^{d^1+b^3}_x\partial^{\delta^1+2q^1}_\xi F_\z]
(x-\tau y,\xi-\tau\eta)\big| \leq c\;\!\<\xi-\tau\eta\>^{m-\rho(|\alpha|+|b|+|\delta^1|+2|q^1|)}
\end{equation*}
with $c$ independent of $x,y,\xi,\eta,\tau$ and $\z$.

We now study the dependence on $\z$ of
$\big|f(x-\tau z,\xi-\tau\zeta)-\z\big|^{-1}$. Clearly, if $\z'\leq \z$, then $\big|f(x-\tau z,\xi-\tau\zeta)-\z'\big|^{-1}
\leq \big|f(x-\tau z,\xi-\tau\zeta)-\z\big|^{-1}$, but this trivial estimate is going to be necessary but not sufficient.
Then, by using the ellipticity of $f$, one obtains that there exist $\kappa,\kappa_1,\kappa_2 \in \R_+$, depending only on $f$,
such that for $|\z|$ large enough one has
$\big|f(x-\tau z,\xi-\tau\zeta)-\z\big|^{-1}\leq \kappa_1 \<\tau\zeta\>^m\big(\kappa_2\<\xi\>^m+|\z|-\kappa\big)^{-1}$.
One can then take into account the inequality $\kappa_2\<\xi \>^m+|\z|-\kappa\geq \mu^{1/\mu}\;\!
(\nu \kappa_2)^{1/\nu}\;\!(|\z|-\kappa)^{1/\mu}\;\!\<\xi\>^{m/\nu}$, valid for any
$\mu,\nu\geq 1$ with $\mu^{-1}+\nu^{-1}=1$, and obtains
\begin{equation*}
\big|f(x-\tau z,\xi-\tau\zeta)-\z\big|^{-1} \leq c \;\!(|\z|-\kappa)^{-1/\mu}\;\!\<\zeta\>^m
\;\!\<\xi\>^{-m/\nu}
\end{equation*}
with $c$ dependent only on $f,\mu$ and $\nu$.

Now, recall that $G_\z=[\partial^{b+\alpha'}_x \partial^{a+\beta}_\xi (f-\z)^{-1}]$. Similarly to $F_\z$ two
distinct situations have to be considered: If $b=\alpha'=a=\beta=d^2=c^3=\delta^2=q^2=0$, then
\begin{equation*}
\big|[\partial^{d^2+c^3}_x\partial^{\delta^2+2q^2}_\xi G_\z](x-\tau z,\xi-\tau\zeta)\big| \equiv
\big|f(x-\tau z,\xi-\tau\zeta)-\z\big|^{-1}\ .
\end{equation*}
But if any of these multi-indices is non-null, then it is not difficult to obtain that
\begin{equation}\label{case2}
\big|[\partial^{d^2+c^3}_x\partial^{\delta^2+2q^2}_\xi G_\z](x-\tau z,\xi-\tau\zeta)\big| \leq
d\;\!|f(x-\tau z,\xi-\tau\zeta)-\z|^{-2}\;\!\<\xi-\tau\zeta\>^{m-\rho(|a|+|\beta|+|\delta^2|+2|q^2|)}
\end{equation}
with $d$ independent of $x,z,\xi,\zeta,\tau$ and $\z$.

So, let us first consider the simple situation, {\it i.e.}~at least one of the multi-indices $a,\beta',\alpha,b,d^1,b^3,
\delta^1,q^1$ is non-null. Then, by taking into account the above estimates, the explicit form of $\sb$ and
Lemma \ref{sticloasa}, one obtains that for any $\tau \in [0,1]$ the following inequalities hold:
\begin{eqnarray*}
\big|L^{\tau, \z,\delta^1, \delta^2}_{p, q, d^0, d^1, d^2}(X,Y,Z)\big|&\leq& C^{\delta^1 \delta^2}_{pq
d^0 d^1 d^2}\; \<\eta\>^{-2p}\<\zeta\>^{-2p}\<y\>^{-2q}\<z\>^{-2q}\
\big(\<y\>+\<z\>\big)^{|d|+|\alpha|+|\beta|+4p}\ \cdot \\
&&\cdot \;
\<\xi-\tau\eta\>^{m-\rho(|\alpha|+|b|+|\delta^1|+2|q^1|)} \;\!
|f(x-\tau z,\xi-\tau\zeta)-\z|^{-1}\;\!\<\xi-\tau\zeta\>^{-\rho(|a|+|\beta|+|\delta^2|+2|q^2|)}
\\
&\leq& D^{\delta^1 \delta^2}_{pq
d^0 d^1 d^2}\;\!(|\z|-\kappa)^{-1/\mu}\;\!
\<\eta\>^{-2p+|m-\rho(|\alpha|+|b|+|\delta^1|)|}\;\!
\<\zeta\>^{-2p+m+\rho(|a|+|\beta|+|\delta^2|)}\ \cdot \\
&&\cdot \;\<y\>^{-2q+|\alpha|+|\beta|+4p+|d|}\;\!
\<z\>^{-2q+|\alpha|+|\beta|+4p+|d|}\;\!
\<\xi\>^{m(1-1/\nu)-\rho(1+|\delta|)},
\end{eqnarray*}
where the trivial inequality mentioned above has been used once for the first inequality.

In the critical case, {\it i.e.}~$a=\beta'=\alpha=b=d^1=b^3=\delta^1=q^1=0$,
one has $|\beta|=1$ because of the definition of $R_\z$ given in \eqref{defofR}.
Thus, we are not in the exceptional case for $G_\z$ and \eqref{case2} always holds.
So, let us consider the following inequalities:
\begin{eqnarray*}
\Big|\frac{f(x-\tau y,\xi-\tau\eta)-\z}{f(x-\tau z,\xi-\tau\zeta)-\z}\Big|
&\leq& 1\!\! + \Big|\sum_{j=1}^n \tau(z_j-y_j)\frac{\int_0^1 \de s [\partial_{x_j} f]
\big(x-\tau z+s\tau(z-y),\xi-\tau \zeta+s\tau(\zeta-\eta)\big)}{f(x-\tau z,\xi-\tau\zeta)-\z}\Big| \\
&&+\Big|\sum_{j=1}^n \tau(\zeta_j-\eta_j)\frac{\int_0^1 \de s [\partial_{\xi_j} f]
\big(x-\tau z+s\tau(z-y),\xi-\tau \zeta+s\tau(\zeta-\eta)\big)}{f(x-\tau z,\xi-\tau\zeta)-\z}\Big|\\
&\leq&1+c\;\!\big|f(x-\tau z,\xi-\tau\zeta)-\z\big|^{-1}
\;\!\<y\>\;\!\<z\>\;\!\<\eta\>^{m+1-\rho}\;\!\<\zeta\>^{m+1-\rho}\;\!\<\xi\>^m \\
&\leq& 1+d\;\!\<y\>\;\!\<z\>\;\!\<\eta\>^{m+1-\rho}\;\!\<\zeta\>^{2m+1-\rho}
\end{eqnarray*}
with $c,d$ independent of all variables and of $\z$. By using these inequalities
one obtains in the critical case:
\begin{eqnarray*}
\big|L^{\tau, \z,\delta^1, \delta^2}_{p, q, d^0, d^1, d^2}(X,Y,Z)\big|&\leq& C^{\delta^1 \delta^2}_{pq
d^0 d^1 d^2}\; \<\eta\>^{-2p}\<\zeta\>^{-2p}\<y\>^{-2q}\<z\>^{-2q}\
\big(\<y\>+\<z\>\big)^{|d|+1+4p}\ \cdot \\
&\cdot &
\Big|\frac{f(x-\tau y,\xi-\tau\eta)-\z}{f(x-\tau z,\xi-\tau\zeta)-\z}\Big|
|f(x-\tau z,\xi-\tau\zeta)-\z|^{-1}\;\!\<\xi-\tau\zeta\>^{m-\rho(1+|\delta^2|+2|q^2|)}
\\
&\leq& D^{\delta^1 \delta^2}_{pq d^0 d^1 d^2}\;\!(|\z|-\kappa)^{-1/\mu}
\<\eta\>^{-2p}\<\zeta\>^{-2p}\<y\>^{-2q+1+4p+|d|}\<z\>^{-2q+1+4p+|d|} \ \cdot \\
&\cdot & \big[1+d\;\!\<y\>\;\!\<z\>\;\!\<\eta\>^{m+1-\rho}\;\!\<\zeta\>^{2m+1-\rho}\big]
\;\!\<\zeta\>^{m+|m-\rho(1+|\delta|)|}
\;\!\<\xi\>^{m(1-1/\nu)-\rho(1+|\delta|)}.
\end{eqnarray*}
Then, it only remains to insert these estimates for
$L^{\tau, \z,\delta^1, \delta^2}_{p, q, d^0, d^1, d^2}$ into the expression of
$R_\z$, and to observe that by choosing $p$ large enough, one gets absolute
integrability in $\eta$ and $\zeta$. A subsequent choice of $q$
also ensures integrability in $y$ and $z$.

We are now in a position to obtain estimates for \eqref{zdependent}. By summing the
contributions in the critical case and in the regular one, we obtain:
\begin{equation*}
\<\xi\>^{\rho(|\delta|-|d|)}\;\!|\partial_x^d \partial_{\xi}^{\delta} R_\z(x,\xi)| \leq
c \;\!(|\z|-\kappa)^{-1/\mu}\;\!\<\xi\>^{m(1-1/\nu)-\rho(1+|d|)}
\end{equation*}
with $c$ independent of $\z$, $x$ and $\xi$. Then, by choosing $\nu$ close enough to $1$
such that $m(1-1/\nu)-\rho<0$, the expression decrease as $|\z|$
increases. Thus, for $|\z|$ big enough, $\| R_\z\|_B$ is strictly
less than $1$ and the Neumann series in then convergent. It follows that
$(f-\z)^{(-1)_B}$ belongs to $\BBA$ for any $\z \in \R_-$ with $|\z|$
large enough.
Finally, by an argument similar to the one proposed in the proof of Theorem 1.8
of \cite{MPR2}, one can extend this result to any $\z \in \C\setminus \R$
and show that the resolvent equation is satisfied. Then, the general argument
already quoted in the proof of Theorem \ref{afileisn} allows us to conclude.

\end{document}